\documentclass[12pt,english,onecolumn,draftcls]{IEEEtran}
\usepackage[T1]{fontenc}
\usepackage[latin9]{inputenc}
\usepackage{float}
\usepackage{bm}
\usepackage{amsmath}
\usepackage{amsthm}
\usepackage{amssymb}
\usepackage{graphicx}

\makeatletter

\providecommand{\tabularnewline}{\\}
\floatstyle{ruled}
\newfloat{algorithm}{tbp}{loa}
\providecommand{\algorithmname}{Algorithm}
\floatname{algorithm}{\protect\algorithmname}

\theoremstyle{plain}
\newtheorem{thm}{\protect\theoremname}
\theoremstyle{plain}
\newtheorem{prop}{\protect\propositionname}
\theoremstyle{remark}
\newtheorem{rem}{\protect\remarkname}
\theoremstyle{definition}
\newtheorem{defn}{\protect\definitionname}
\theoremstyle{plain}
\newtheorem{lem}{\protect\lemmaname}

\makeatother

\usepackage{babel}
\providecommand{\definitionname}{Definition}
\providecommand{\lemmaname}{Lemma}
\providecommand{\propositionname}{Proposition}
\providecommand{\remarkname}{Remark}
\providecommand{\theoremname}{Theorem}

\begin{document}

\title{Mixed-Timescale Online PHY Caching for Dual-Mode MIMO Cooperative
Networks}

\author{{\normalsize{}An Liu$^{1}$, Vincent Lau$^{2}$, Wenchao Ding$^{2}$
and Edmund Yeh$^{3}$\\$^{1}$College of Information Science and
Electronic Engineering, Zhejiang University\\$^{2}$Department of
ECE, Hong Kong University of Science and Technology\\$^{3}$Department
of Electrical and Computer Engineering, Northeastern University\vspace{-0.3in}
}}
\maketitle
\begin{abstract}
Recently, physical layer (PHY) caching has been proposed to exploit
the \emph{dynamic side information} induced by caches at base stations
(BSs) to support Coordinated Multi-Point (CoMP) and achieve huge degrees
of freedom (DoF) gains. Due to the limited cache storage capacity,
the performance of PHY caching depends heavily on the cache content
placement algorithm. In existing algorithms, the cache content placement
is adaptive to the long-term popularity distribution in an offline
manner. We propose an online PHY caching framework which adapts the
cache content placement to \textit{microscopic spatial and temporary
popularity variations} to fully exploit the benefits of PHY caching.
Specifically, the joint optimization of online cache content placement
and content delivery is formulated as a mixed-timescale drift minimization
problem to increase the CoMP opportunity and reduce the cache content
placement cost. We propose a low-complexity algorithm to obtain a
throughput-optimal solution. Moreover, we provide a closed-form characterization
of the maximum sum DoF in the stability region and study the impact
of key system parameters on the stability region. Simulations show
that the proposed online PHY caching framework achieves large gain
over existing solutions.
\end{abstract}


\section{Introduction}

Many recent works have shown that wireless network performance can
be substantially improved by exploiting caching in content-centric
wireless networks \cite{Tao_tvt2015_caching,Tao_AC2016_caching}.
The early works \cite{golrezaei2012femtocaching,dai2012collaborative,ji2013fundamental}
focus on exploiting caching to reduce the backhaul loading. However,
the performance of wireless networks is fundamentally limited by interference,
and these existing works fail to address this issue. In \cite{liu2013mixed,liu2014cache},
cache-enabled opportunistic CoMP (PHY caching) is proposed to mitigate
interference and improve the spectral efficiency of the physical layer
(PHY) in wireless networks with limited backhaul. Beside the common
benefits of caching in fixed line networks, DoF gains can be achieved
by utilizing a fundamental \textit{cache-induced PHY topology change}.
Specifically, when users' requested content exists in the cache of
several BSs, the BS caches induce \emph{dynamic side information},
which can be used to cooperatively transmit the requested packets
to the users, thus achieving huge DoF gains without requiring high-capacity
backhaul links. In practice, the cache storage capacity is limited,
and hence, the cache content placement algorithm plays a key role
in determining the performance of PHY caching schemes. The existing
algorithms can be classified into two types.

\textbf{Offline cache content placement:} Offline cache content placement
is adaptive to the long-term popularity distribution in an offline
manner. Once the cache content placement phase has finished, the cached
content cannot be changed during the content delivery phase. In \cite{liu2013mixed,liu2014cache},
mixed-timescale optimizations of short-term MIMO precoding and long-term
cache content placement are studied to support real-time video-on-demand
applications. However, these offline caching algorithms cannot capture
the \textit{microscopic spatial and temporary popularity variations}.

\textbf{Online cache content placement:} Online cache content placement
is dynamically adaptive to microscopic spatial and temporary popularity
variations during the content delivery phase. Compared to its offline
counterpart, it has more refined control over the limited cache resources
and thus can potentially achieve a better performance. In \cite{abedini2014content},
online cache placement and request scheduling are studied to support
elastic and inelastic traffic in wireless networks. In \cite{yeh2014vip},
a framework for joint online forwarding and caching is proposed within
the context of Name Data Networks (NDNs). The throughput-optimal
solution in \cite{yeh2014vip} is to cache the content with the longest
request queue at each node. In other words, the caching priority is
based solely on the local popularity of content. However, the solution
in \cite{yeh2014vip} cannot be extended easily to PHY caching with
cache-induced CoMP. This is because the local popularity at different
BSs may vary widely, yielding a low cooperation opportunity. As such,
a reasonable online caching policy should strike a delicate balance
between local popularity at individual BSs and the cooperation opportunity
among cooperative BSs.

In this paper, we propose an online PHY caching framework to fully
exploit the benefits of cache-induced opportunistic CoMP. The main
contributions are summarized as follows.
\begin{itemize}
\item \textbf{Online PHY caching with cache content placement cost: }In
\cite{amble2011content,yeh2014vip}, the cache content placement cost
is ignored, and hence, the cache content placement policy depends
only on the current cache state. However, for practical consideration,
it is important to model the cache content placement cost in the system
because cache content update will cause communication overhead in
the network. In this case, the cache content placement policy should
depend on both the previous and the current cache states. As a result,
both the algorithm design and throughput optimality analysis are more
complicated because they involve a history-dependent policy, as explained
below.
\item \textbf{Mixed-timescale optimization of online PHY caching and content
delivery: }We apply the Lyapunov optimization framework \cite{neely2006energy,neely2010stochastic}
to address the joint optimization of online PHY caching and content
delivery. In our design, the cache content placement is updated at
a slower timescale than the other control variables to reduce the
cache content placement cost. With such mixed-timescale control variables,
the algorithm design is based on minimizing a $T$-step VIP-drift-plus-penalty
function, which contains both the previous and the current cache states.
As such, the conventional Lyapunov optimization algorithms based on
minimizing a 1-step drift-plus-penalty function over single-timescale
control variables cannot be applied. By exploiting the specific structure
of the problem, we propose a low-complexity mixed-timescale optimization
algorithm and establish its throughput optimality. Due to the history-dependent
policy and the mixed-timescale design, the throughput optimality analysis
cannot directly follow the routine of conventional Lyapunov drift
plus penalty theory in \cite{neely2006energy,neely2010stochastic}.
To over come this challenge, we first introduce the concept of conditional
flow balance constraint for a given cache state, and define the network
stability region under conditional flow balance. Then we establish
the throughput optimality by introducing a FRAME policy as a bridge
to connect the proposed solution and the optimal random policy.
\item \textbf{Closed-form characterization of the stability region:} We
provide a simple characterization of the stability region (in terms
of DoF), incorporating the effect of cooperative caching so as to
study the impact of key system parameters on the stability region. 
\end{itemize}

The online PHY caching has been proposed in the conference version
\cite{Liu_GC2017_VIPcaching}, but without the analysis of throughput
optimality and stability region. The rest of the paper is organized
as follows: In Section \ref{sec:System-Model}, we introduce the system
model. In Section \ref{sec:Mixed-Timescale-Online}, we elaborate
the proposed online PHY caching and content delivery schemes. In Section
\ref{sec:Dual-Mode-VIP-based-Resource-Con}, we propose a dual-mode-VIP-based
resource control framework for wireless NDNs with dual-mode PHY. In
Section \ref{sec:Throughput-Optimality-Analysis}, we establish the
throughput optimality of the proposed resource control algorithm.
In Section \ref{sec:Characterization-of-the}, we characterize the
stability region of wireless NDNs with a dual-mode PHY. Simulations
are presented in Section \ref{sec:Simulation-Results}, and conclusions
are given in Section \ref{sec:Conclusion}.

\section{System Model\label{sec:System-Model}}

\subsection{Network Architecture}

\begin{figure}
\begin{centering}
\includegraphics[width=85mm]{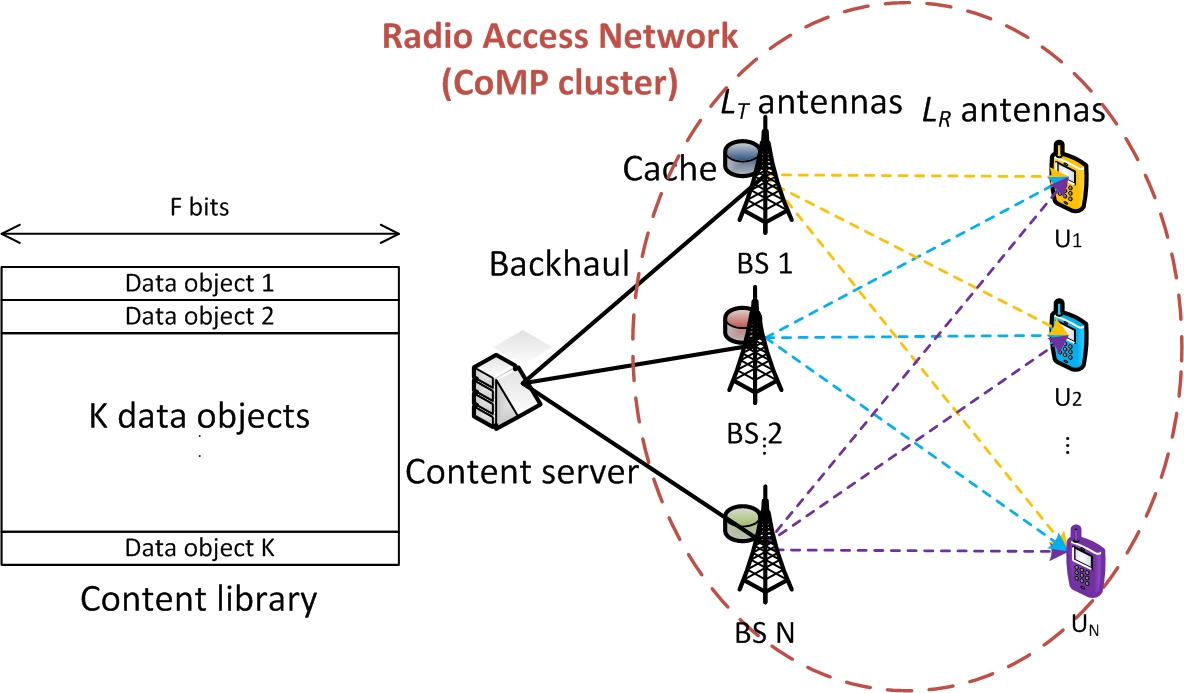}
\par\end{centering}
\caption{\label{fig:Architecture}{\small{}Architecture of cached MIMO interference
networks.}}
\end{figure}

Consider a cached MIMO interference network with $N$ BS-user pairs,
as illustrated in Fig. \ref{fig:Architecture}. Each BS has $L_{T}$
antennas and each user has $L_{R}$ antennas. Each BS has transmit
power $P$ and a cache of $L_{C}F$ bits. There is a content server
providing a content library $\mathcal{K}$ that contains $K$ data
objects, where each data object has $F$ bits. The users request data
objects from the content server via a radio access network (RAN).
Each BS in the RAN is connected to the content server via a backhaul.
The content server also serves as a central control node responsible
for resource control of all the BSs. Such a cluster architecture appears
in practical LTE networks. 

For convenience, let $\mathcal{B}$ denote the set of BSs, $\mathcal{U}$
denote the set of users, $\mathcal{M}=\mathcal{B}\cup\mathcal{U}$
denote the set of all nodes, and $g$ denote the content server. The
serving BS of user $j\in\mathcal{U}$ is denoted as $n_{j}$ and the
associated user of BS $n\in\mathcal{B}$ is denoted as $j_{n}$. A
user always sends its data object request to its serving BS. However,
it may receive the requested data object from only the serving BS
or from all BSs, depending on the PHY mode, as will be elaborated
in the next subsection.

\begin{figure}
\centering{}%
\begin{minipage}[t]{0.45\textwidth}%
\begin{center}
\includegraphics[width=80mm]{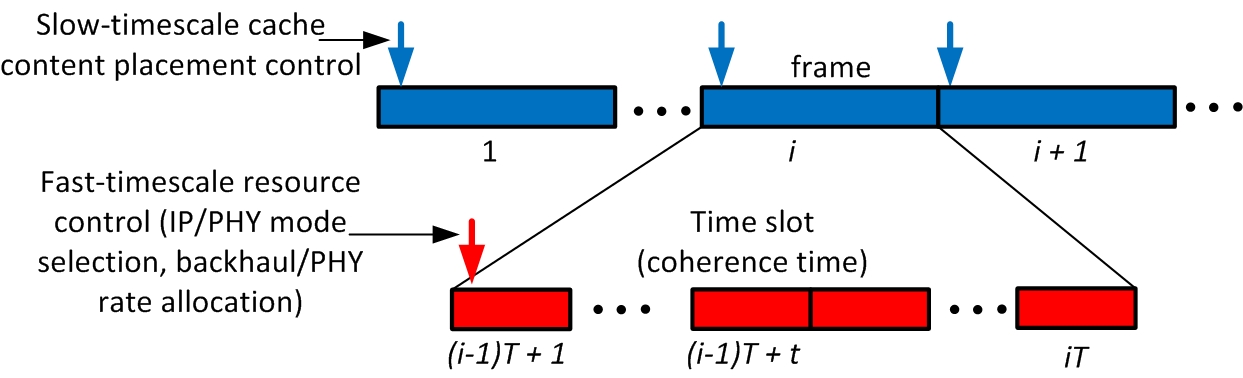}
\par\end{center}
\begin{center}
\caption{\label{fig:frame}{\small{}Illustration of the time slot and frame.}}
\par\end{center}%
\end{minipage}\hfill{}%
\begin{minipage}[t]{0.45\textwidth}%
\begin{center}
\includegraphics[width=80mm]{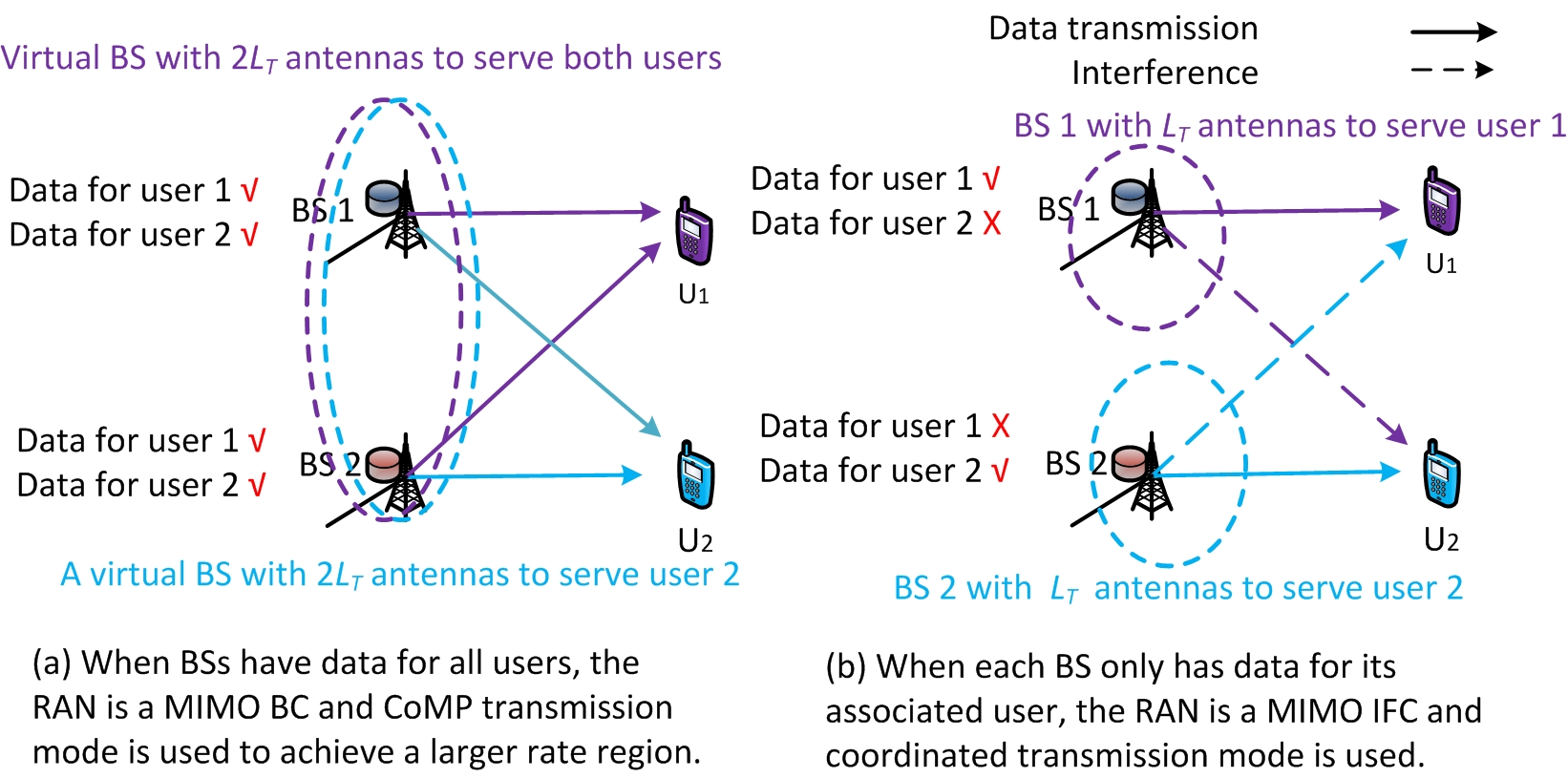}
\par\end{center}
\begin{center}
\caption{\label{fig:cooperative-caching}{\small{}Illustration of the dual-mode
PHY.}}
\par\end{center}%
\end{minipage}
\end{figure}

Time is partitioned into frames indexed by $i$, and each frame consists
of $T$ time slots indexed by $t$, as illustrated in Fig. \ref{fig:frame}.
Each time slot corresponds to a channel coherence time and the fast-timescale
resource control variables (mode selection and rate allocation) are
updated at the beginning of each time slot. On the other hand, the
slow-timescale resource control variables (cache content placement
control) are updated at the beginning of each frame. Unless otherwise
specified, $t$ is used to index a time slot in frame $i$, i.e.,
$t\in\left[1+(i-1)T,iT\right]$ and $i=\left\lceil \frac{t}{T}\right\rceil $.

\subsection{Dual-Mode Physical Layer Model}

We assume that the content server has knowledge of the global channel
state information (CSI) $\boldsymbol{H}\left(t\right)=\left\{ \boldsymbol{H}_{jn}\left(t\right),\forall j\in\mathcal{U},n\in\mathcal{B}\right\} $
of the RAN, where $\boldsymbol{H}_{jn}\left(t\right)\in\mathbb{C}^{L_{R}\times L_{T}}$
is the channel matrix between the $n$-th BS and the $j$-th user
at the $t$-th time slot, and $\boldsymbol{H}\left(t\right)$ is quasi-static
within a time slot and i.i.d. between time slots. The time index $t$
in $\boldsymbol{H}\left(t\right)$ will be omitted when there is no
ambiguity. There are two PHY modes, as elaborated below. 

\textbf{CoMP transmission mode (PHY mode A):} In this mode, the BSs
can form a virtual transmitter and cooperatively send some data to
all users. In this case, the RAN is a virtual MIMO broadcast channel
(BC), as illustrated in Fig. \ref{fig:cooperative-caching}-(a). Specifically,
let $B_{j}\left(t\right)$ denote the data scheduled for delivery
to user $j$ at time slot $t$. In CoMP mode, $B_{j}\left(t\right)$
must be available at all BSs. The amount of scheduled data $\left|B_{j}\left(t\right)\right|$
(measured by the number of data objects) is limited by the data rate
of the CoMP mode PHY, i.e., $\left|B_{j}\left(t\right)\right|=c_{j}^{A}\left(t\right)$,
where $c_{j}^{A}\left(t\right)$ (data objects/slot) is the data rate
of user $j$ under CoMP mode with CSI $\boldsymbol{H}\left(t\right)$.
To achieve the data rate $c_{j}^{A}\left(t\right)$, a coding scheme
$\beta_{nj}^{A}:\left\{ 0,1\right\} ^{c_{j}^{A}\left(t\right)F}\rightarrow\mathbb{C}^{L_{T}\times N_{c}}$
is applied at BS $n,\forall n\in\mathcal{B}$ for the data $B_{j}\left(t\right)$
requested by user $j$, which maps $B_{j}\left(t\right)$ to a codeword:
$\boldsymbol{X}_{nj}^{A}\left(t\right)=\beta_{nj}^{A}\left(B_{j}\left(t\right)\right)\in\mathbb{C}^{L_{T}\times N_{c}}$,
where $N_{c}$ is the number of data symbol vectors per time slot.
The signal $\boldsymbol{X}_{n}^{A}\left(t\right)$ transmitted from
BS $n$ in the $t$-th time slot is a superposition of the codewords
for all users: $\boldsymbol{X}_{n}^{A}\left(t\right)=\sum_{j\in\mathcal{U}}\boldsymbol{X}_{nj}^{A}\left(t\right)\in\mathbb{C}^{L_{T}\times N_{c}}$,
where each column vector in $\boldsymbol{X}_{n}^{A}\left(t\right)$
is transmitted from the $L_{T}$ antennas during a symbol period in
time slot $t$. Moreover, $\boldsymbol{X}_{n}^{A}\left(t\right)$
satisfies a power constraint $\mathrm{Tr}\left(\boldsymbol{X}_{n}^{A}\left(t\right)\left(\boldsymbol{X}_{n}^{A}\left(t\right)\right)^{H}\right)/N_{c}\leq P$.
The received signal at user $j$ in the $t$-th time slot is
\begin{equation}
\boldsymbol{Y}_{j}\left(t\right)=\sum_{n\in\mathcal{B}}\boldsymbol{H}_{jn}\left(t\right)\boldsymbol{X}_{n}^{A}\left(t\right)+\boldsymbol{Z}_{j}\left(t\right)\in\mathbb{C}^{L_{R}\times N_{c}},\label{eq:revCoMP}
\end{equation}
where $\boldsymbol{Z}_{j}\left(t\right)$ is the additive white Gaussian
noise (AWGN). The rate $c_{j}^{A}\left(t\right)$ is achievable at
the $t$-th time slot if user $j$ can decode the scheduled data $B_{j}\left(t\right)$
with vanishing error probability as $N_{c}\rightarrow\infty$. The
set of achievable rate vectors $\boldsymbol{c}^{A}\left(t\right)=\left[c_{j}^{A}\left(t\right)\right]_{j\in\mathcal{U}}\in\mathbb{R}_{+}^{N}$
forms the capacity region $C^{A}\left(\boldsymbol{H}\left(t\right)\right)\in\mathbb{R}_{+}^{N}$
under the CoMP mode. 

\textbf{Coordinated transmission mode (PHY mode B):} In this case,
user $j$ can only be served by the serving BS $n_{j}$ and the RAN
is a MIMO interference channel (IFC), as illustrated in Fig. \ref{fig:cooperative-caching}-(b).
Similarly, we have $\left|B_{j}\left(t\right)\right|=c_{j}^{B}\left(t\right)$,
where $c_{j}^{B}\left(t\right)$ (data objects/slot) is the data rate
of user $j$ under coordinated mode with CSI $\boldsymbol{H}\left(t\right)$.
To achieve the data rate $c_{j}^{B}\left(t\right)$, a coding scheme
$\beta_{j}^{B}:\left\{ 0,1\right\} ^{c_{j}^{B}\left(t\right)F}\rightarrow\mathbb{C}^{L_{T}\times N_{c}}$
is applied at BS $j_{n}$ to map $B_{j}\left(t\right)$ to a codeword:
$\boldsymbol{X}_{j}^{B}\left(t\right)=\beta_{j}^{B}\left(B_{j}\left(t\right)\right)\in\mathbb{C}^{L_{T}\times N_{c}}$,
where $\boldsymbol{X}_{j}^{B}\left(t\right)$ satisfies a power constraint
$\mathrm{Tr}\left(\boldsymbol{X}_{j}^{B}\left(t\right)\left(\boldsymbol{X}_{j}^{B}\left(t\right)\right)^{H}\right)/N_{c}\leq P$.
The received signal at user $j$ in the $t$-th time slot is
\begin{equation}
\boldsymbol{Y}_{j}\left(t\right)=\sum_{j^{'}\in\mathcal{K}}\boldsymbol{H}_{jn_{j^{'}}}\left(t\right)\boldsymbol{X}_{j^{'}}^{B}\left(t\right)+\boldsymbol{Z}_{j}\left(t\right)\in\mathbb{C}^{L_{R}\times N_{c}}.\label{eq:revCoor}
\end{equation}
The rate $c_{j}^{B}\left(t\right)$ at the $t$-th time slot is achievable
if the user $j$ can decode the scheduled data $B_{j}\left(t\right)$
with vanishing error probability as $N_{c}\rightarrow\infty$. The
set of achievable rate vectors $\boldsymbol{c}^{B}\left(t\right)=\left[c_{j}^{B}\left(t\right)\right]_{j\in\mathcal{U}}\in\mathbb{R}_{+}^{N}$
forms the capacity region $C^{B}\left(\boldsymbol{H}\left(t\right)\right)\in\mathbb{R}_{+}^{N}$
under the coordinated transmission mode. 

There exist many CoMP/coordinated transmission schemes (coding schemes
$\left\{ \beta_{nj}^{A}\right\} /\left\{ \beta_{j}^{B}\right\} $).
In this paper, we do not restrict the PHY to be any specific CoMP/coordinated
transmission scheme but consider an abstract PHY model represented
by the capacity regions $C^{A}\left(\boldsymbol{H}\right)$ and $C^{B}\left(\boldsymbol{H}\right)$.
Note that $C^{A}\left(\boldsymbol{H}\right)$ and $C^{B}\left(\boldsymbol{H}\right)$
depend on the transmit power $P$ at each BS. Moreover, we have $C^{B}\left(\boldsymbol{H}\right)\subseteq C^{A}\left(\boldsymbol{H}\right)$
according to the definitions of $C^{A}\left(\boldsymbol{H}\right)$
and $C^{B}\left(\boldsymbol{H}\right)$.

In the following, we use linear precoding to illustrate the dual mode
physical layer. 

\textbf{CoMP Mode} \textbf{under linear precoding: }In this mode,
the $N$ users are served using CoMP linear precoding between the
BSs. The received signal for user $j$ can be expressed as:
\begin{equation}
\boldsymbol{y}_{j}=\widetilde{\boldsymbol{H}}_{j}\boldsymbol{V}_{j}^{A}\boldsymbol{x}_{j}^{A}+\sum_{j^{'}\neq j}\widetilde{\boldsymbol{H}}_{j}\boldsymbol{V}_{j^{'}}^{A}\boldsymbol{x}_{j^{'}}^{A}+\boldsymbol{z}_{j},\label{eq:recCoMPlin}
\end{equation}
where $\widetilde{\boldsymbol{H}}_{j}=\left[\boldsymbol{H}_{j1},...,\boldsymbol{H}_{jN}\right]\in\mathbb{C}^{L_{R}\times NL_{T}}$
is the composite channel matrix between all the BSs and user $j$;
$\boldsymbol{x}_{j}^{A}\in\mathbb{C}^{d_{j}^{A}}\sim\mathcal{CN}\left(0,\boldsymbol{I}\right)$
and $d_{j}^{A}$ are respectively the data vector and the number of
data streams for user $j$; $\boldsymbol{V}_{j}^{A}\in\mathbb{C}^{NL_{T}\times d_{j}^{A}}$
is the composite precoding matrix for user $j$, and $\boldsymbol{z}_{j}$
is the AWGN. For given CSI $\boldsymbol{H}$ and precoding matrices
$\boldsymbol{V}^{A}=\left\{ \boldsymbol{V}_{j}^{A}:\forall j\right\} $,
the data rate (bps) of user $j$ under the CoMP mode is \cite{Liu_TSP13_CacheIFN}
\begin{equation}
c_{j}^{A}\left(\boldsymbol{H},\boldsymbol{V}^{A}\right)=B_{W}\textrm{log}_{2}\left|\boldsymbol{I}+\widetilde{\boldsymbol{H}}_{j}\boldsymbol{V}_{j}^{A}\boldsymbol{V}_{j}^{AH}\widetilde{\boldsymbol{H}}_{j}^{H}\widetilde{\mathbf{\Omega}}_{j}^{-1}\right|,\label{eq:Rk1}
\end{equation}
where $B_{W}$ is the channel bandwidth, and $\widetilde{\mathbf{\Omega}}_{j}=\boldsymbol{I}+\sum_{j^{'}\neq j}\widetilde{\boldsymbol{H}}_{j}\boldsymbol{V}_{j^{'}}^{A}\boldsymbol{V}_{j^{'}}^{H}\widetilde{\boldsymbol{H}}_{j}^{H}$
is the interference-plus-noise covariance matrix of user $j$.

\textbf{Coordinated mode under linear precoding:} In this mode, the
user $j$ can only be served by BS $j$ using coodinated linear precoding.
The received signal for user $j$ can be expressed as:
\begin{equation}
\boldsymbol{y}_{j}=\boldsymbol{H}_{jn_{j}}\boldsymbol{V}_{j}^{B}\boldsymbol{x}_{j}^{B}+\sum_{j^{'}\neq j}\boldsymbol{H}_{jn_{j^{'}}}\boldsymbol{V}_{j^{'}}^{B}\boldsymbol{x}_{j^{'}}^{B}+\boldsymbol{z}_{j},\label{eq:recvCoodlin}
\end{equation}
where $\boldsymbol{x}_{j}^{B}\in\mathbb{C}^{d_{j}^{B}}\sim\mathcal{CN}\left(0,\boldsymbol{I}\right)$
and $d_{j}^{B}$ are respectively the data vector and the number of
data streams for user $j$; and $\boldsymbol{V}_{j}^{B}\in\mathbb{C}^{L_{T}\times d_{j}^{B}}$
is the precoding matrix for user $j$. For given CSI $\boldsymbol{H}$
and precoding matrices $\boldsymbol{V}^{B}=\left\{ \boldsymbol{V}_{j}^{B}:\forall j\right\} $,
the data rate of user $j$ under coordinated mode is \cite{Liu_TSP13_CacheIFN}
\begin{equation}
c_{j}^{B}\left(\boldsymbol{H},\boldsymbol{V}^{B}\right)=B_{W}\textrm{log}_{2}\left|\boldsymbol{I}+\boldsymbol{H}_{jn_{j}}\boldsymbol{V}_{j}^{B}\boldsymbol{V}_{j}^{BH}\boldsymbol{H}_{jn_{j}}^{H}\mathbf{\Omega}_{j}^{-1}\right|,\label{eq:Rk}
\end{equation}
where $\mathbf{\Omega}_{j}=\boldsymbol{I}+\sum_{j^{'}\neq j}\boldsymbol{H}_{jn_{j^{'}}}\boldsymbol{V}_{j^{'}}^{B}\boldsymbol{V}_{j^{'}}^{BH}\boldsymbol{H}_{jn_{j^{'}}}^{H}$
is the interference-plus-noise covariance matrix.

\section{Mixed-Timescale Online PHY Caching and Content Delivery Scheme\label{sec:Mixed-Timescale-Online}}

\subsection{Slow-Timescale Online PHY Caching Scheme}

In the proposed online PHY caching scheme, the cached data objects
at each BS are updated once every\textit{ }frame ($T$ time slots),
as illustrated in Fig. \ref{fig:frame}. Since the local popularity
variations at each BS usually change at a timescale much slower than
the instantaneous CSI (slot interval), in practice, we may choose
$T\gg1$ to reduce the cache content placement cost without losing
the ability to track microscopic spatial and temporary popularity
variations.

Let $s_{n}^{k}(i)\in\left\{ 0,1\right\} $ denote the \textit{cache
state} of data object $k$ at BS $n$, where $s_{n}^{k}(i)=1$ means
that data object $k$ is in the cache of BS $n$ at frame $i$ and
$s_{n}^{k}(i)=0$ means the opposite. The \textit{cache placement
control action} at the beginning of the $i$-th frame is denoted by
$\left\{ p_{n}^{k}(i)\in\left\{ -1,0,1\right\} ,\forall n,k\right\} $,
where $p_{n}^{k}(i)=-1$ and $p_{n}^{k}(i)=1$ mean that the data
object $k$ is removed from and added to the cache of BS $n$ at the
beginning of the $i$-th frame respectively, and $p_{n}^{k}(i)=0$
means that the cache state is unchanged. Note that there is no need
to add an existing data object to the cache or remove a non-existing
data object, i.e.,
\begin{align}
p_{n}^{k}(i) & \neq1,\textrm{when }s_{n}^{k}(i-1)=1.\nonumber \\
p_{n}^{k}(i) & \neq-1,\textrm{when }s_{n}^{k}(i-1)=0.\label{eq:cacheover}
\end{align}
As such, the cache state dynamics is
\begin{equation}
s_{n}^{k}(i)=s_{n}^{k}(i-1)+p_{n}^{k}(i),\forall n\in\mathcal{B},k\in\mathcal{K}.\label{eq:cachedyn}
\end{equation}
The cache placement control action $\left\{ p_{n}^{k}(i)\right\} $
must satisfy the following cache size constraint:
\begin{equation}
\sum_{k\in\mathcal{K}}s_{n}^{k}(i)\leq L_{C},\forall i,\forall n\in\mathcal{B}.\label{eq:cachesizecon}
\end{equation}
Let $\boldsymbol{s}\left(i\right)=\left\{ s_{n}^{k}(i),\forall n\in\mathcal{B},k\in\mathcal{K}\right\} $
denote the \textit{aggregate} \textit{cache state}.\textit{} When
$p_{n}^{k}(i)=1$, BS $n$ needs to obtain data object $k$ from the
backhaul, which induces some \textit{cache content placement cost}.
To accommodate the traffic caused by cache content placement, the
available backhaul capacity $R$ (data objects/slot) at each BS is
divided into a \textit{data sub-channel} and a \textit{control sub-channel}
as $R=R_{c}+R_{d}$, where the data sub-channel with rate $R_{d}$
is used for transmitting the data objects requested by users, and
the control sub-channel is used for transmitting the data objects
induced by the cache placement control and other control signalings.
The cache content placement cost for BS $n$ to cache data object
$k$ at frame $i$ is $\Gamma_{n}^{k}(i)=\gamma\mathbf{1}{}_{\{p_{n}^{k}(i)=1\}}$,
where $\mathbf{1}$ denotes the indication function, and $\gamma$
is the price of fetching one data object using the control sub-channel.
The total cost function at frame $i$ is
\begin{equation}
\Gamma(i)=\sum_{n\in\mathcal{B},k\in\mathcal{K}}\Gamma_{n}^{k}(i)=\sum_{n\in\mathcal{B},k\in\mathcal{K}}\gamma\mathbf{1}{}_{\{p_{n}^{k}(i)=1\}},\label{eq:totalcachepri}
\end{equation}
and the average cache content placement cost is
\begin{equation}
\overline{\Gamma}=\limsup_{J\rightarrow\infty}\frac{1}{J}\sum_{i=1}^{J}\mathbb{E}\left[\Gamma\left(i\right)\right].\label{eq:avgcachecost}
\end{equation}

\subsection{Fast-Timescale Dual-Mode Content Delivery Scheme }

We consider a \textit{dual-mode content delivery scheme} which can
support the CoMP to enhance the capacity of the RAN. Specifically,
each data object is divided into $D$ data chunks, and each data chunk
is allocated a unique ID. The content delivery operates at the level
of data chunks using two types of packets: \textit{Interest Packets}
(IPs) and \textit{Data Packet}s (DPs). To request a data chunk, a
user sends out an IP, which carries the ID of the data chunk, to the
content server via its serving BS. Hence, a request for a data object
consists of a sequence of IPs which request all the data chunks of
the object. 

For each IP from user $j$, the content server determines its mode
(\textit{coordinated mode IP or CoMP mode IP}) according to an \textit{IP
mode selection} policy that will be elaborated in Section \ref{subsec:Virtual-to-Actual-Control-Policy}.
If it is marked as a CoMP Mode IP, the corresponding DP will be delivered
to all BSs and stored in the \textit{CoMP mode data buffer} at each
BS. If it is marked as a coordinated mode IP, the corresponding DP
will be delivered to the serving BS $n_{j}$ only, and stored in a
\textit{coordinated mode data buffer} at BS $n_{j}$. At each time
slot, the content server also needs to determine the PHY mode $M_{a}\left(t\right)\in\left\{ 0,1\right\} $
according to the \textit{PHY mode selection} policy elaborated in
Section \ref{subsec:Virtual-to-Actual-Control-Policy}. If $M_{a}\left(t\right)=0$
($M_{a}\left(t\right)=1$), the BSs will employ the coordinated (CoMP)
mode to transmit some data from the coordinated (CoMP) mode data buffers
to the users. 

The dual-mode content delivery has four components.

\textbf{Component 1 (IP mode selection at content server):} Let $Da_{j}^{k}(t)$
denote the number of IPs of data object $k$ received by the content
server from user $j$ at time slot $t$, where $a_{j}^{k}(t)$ can
be interpreted as the instantaneous arrival rate of IPs in the unit
of data object/slot since each data object corresponds to $D$ IPs.
The content server will mark all these $Da_{j}^{k}(t)$ IPs using
the same \textit{IP mode}, denoted by $m_{j}^{k}\left(t\right)\in\left\{ 0,1\right\} $. 

\textbf{Component 2 (coordinated mode DPs' delivery to each BS):}
If $m_{j}^{k}\left(t\right)=0$, the corresponding $Da_{j}^{k}(t)$
DPs are called \textit{coordinated mode DPs}, which will be delivered
to the serving BS $n_{j}$ only. Specifically, if BS $n_{j}$ has
data object $k$ in the local cache ($s_{n_{j}}^{k}\left(i\right)=1$),
it creates $Da_{j}^{k}(t)$ DPs containing the requested data chunks
indicated by the $Da_{j}^{k}(t)$ IPs. Otherwise, the content server
will create the requested $Da_{j}^{k}(t)$ DPs and store them in the
$n_{j}$-th data buffer with queue length $Q_{gn_{j}}$ at the content
server. These will be sent to BS $n_{j}$ via backhaul when they become
the head-of-the-queue DPs. In both cases, after obtaining the $Da_{j}^{k}(t)$
DPs, BS $n_{j}$ will store them in the coordinated mode data buffer
$Q_{n_{j}j}^{B}$. 

\textbf{Component 3 (CoMP mode DPs' delivery to each BS):} If $m_{j}^{k}\left(t\right)=1$,
the corresponding $Da_{j}^{k}(t)$ DPs are called \textit{CoMP mode
DPs}, which will be delivered to all BSs. For any $n\in\mathcal{B}$,
if BS $n$ has data object $k$ in the local cache ($s_{n}^{k}\left(i\right)=1$),
it creates the requested $Da_{j}^{k}(t)$ DPs. Otherwise, the content
server will create the requested $Da_{j}^{k}(t)$ DPs and store them
in the $n$-th data buffer $Q_{gn}$ at the content server. These
will be sent to BS $n$ via backhaul when they become the head-of-the-queue
DPs. In both cases, after obtaining the $Da_{j}^{k}(t)$ DPs, BS $n$
will store the $Da_{j}^{k}(t)$ DPs in the $j$-th CoMP mode data
buffer $Q_{nj}^{A}$. 

\textbf{Component 4 (PHY mode determination at content server):} The
content server determines the \textit{PHY mode} $M_{a}\left(t\right)$.
If $M_{a}\left(t\right)=0$, coordinated transmission mode will be
used to send the data in $Q_{nj_{n}}^{B}$ to user $j_{n}$, $\forall j_{n}\in\mathcal{U}$,
at rate $c_{j_{n}}^{B}\left(t\right)$ (data objects/slot). If $M_{a}\left(t\right)=1$,
CoMP transmission mode will be used to send the data in $Q_{nj}^{A}$
to user $j$, $\forall j\in\mathcal{U}$, at rate $c_{j}^{A}\left(t\right)$.
Let $\boldsymbol{c}^{B}\left(t\right)=\left[c_{j}^{B}\left(t\right)\right]_{j\in\mathcal{U}}\in\mathbb{R}_{+}^{N}$
and $\boldsymbol{c}^{A}\left(t\right)=\left[c_{j}^{A}\left(t\right)\right]_{j\in\mathcal{U}}\in\mathbb{R}_{+}^{N}$
denote the \textit{PHY rate allocation} for the coordinated and CoMP
modes respectively.

\subsection{Data Packet Queue Dynamics}

The dynamics of the queue $Q_{gn}$ at the content server is
\begin{equation}
Q_{gn}\left(t+1\right)=\left(Q_{gn}(t)-c_{ng}(t)\right)^{+}+b_{n}(t),\label{eq:DPQD}
\end{equation}
where {\small{}$b_{n}(t)=\sum_{k\in\mathcal{K}}\overline{s}_{n}^{k}\left(i\right)\left(\sum_{j\in\mathcal{U}}m_{j}^{k}\left(t\right)a_{j}^{k}(t)+\overline{m}_{j_{n}}^{k}\left(t\right)a_{j_{n}}^{k}(t)\right)$}
is the arrival rate of $Q_{gn}\left(t\right)$, $\overline{s}_{n}^{k}\left(i\right)=\boldsymbol{1}_{s_{n}^{k}\left(i\right)=0}$,
$\overline{m}_{j}^{k}\left(t\right)=\boldsymbol{1}_{m_{j}^{k}\left(t\right)=0}$,
and $c_{ng}(t)$ is the allocated transmission rate of DPs from the
content server to BS $n$ during time slot $t$. Let $Db_{nj_{n}}^{B}(t)$
($Db_{nj}^{A}(t)$) denote the number of coordinated mode DPs of user
$j_{n}$ (CoMP mode IPs of user $j$) obtained at BS $n$ from either
the backhaul or the local cache. The dynamics of the queues at BS
$n$ are
\begin{align}
Q_{nj_{n}}^{B}\left(t+1\right) & =\left(Q_{nj_{n}}^{B}\left(t\right)-\overline{M}_{a}\left(t\right)c_{j_{n}}^{B}\left(t\right)\right)^{+}+b_{nj_{n}}^{B}(t),\nonumber \\
Q_{nj}^{A}\left(t+1\right) & =Q_{nj}^{A}\left(t\right)-\min\left(M_{a}\left(t\right)c_{j}^{A}\left(t\right),\check{Q}_{j}^{A}\left(t\right)\right)+b_{nj}^{A}(t),\label{eq:DPQDBS}
\end{align}
where $\overline{M}_{a}\left(t\right)=\boldsymbol{1}_{M_{a}\left(t\right)=0}$
and $\check{Q}_{j}^{A}\left(t\right)=\min_{n}Q_{nj}^{A}\left(t\right)$.
Note that the queue length is measured using the number of data objects
and the arrival rate is measured in data objects/slot because the
proposed resource control framework operates at the data object level. 

\section{Dual-Mode-VIP-based Resource Control\label{sec:Dual-Mode-VIP-based-Resource-Con}}

In the proposed scheme, the slow-timescale cache content placement
policy $\left\{ p_{n}^{k}(i)\right\} $ is adaptive to the local popularity
information at each node. The fast-timescale controls include the
IP mode selection $\left\{ m_{j}^{k}\left(t\right)\right\} $, backhaul
rate allocation $\left\{ c_{ng}(t)\right\} $, PHY mode selection
$M_{a}\left(t\right)$, and PHY rate allocation $\left\{ \boldsymbol{c}^{A}\left(t\right),\boldsymbol{c}^{B}\left(t\right)\right\} $,
which are adaptive to the cache state $\left\{ s_{n}^{k}(i)\right\} $
and global CSI $\boldsymbol{H}\left(t\right)$. However, the local
popularity information is unavailable in the actual plane (network)
due to interest collapsing and suppression, which refers to the operation
that multiple unsatisfied requests (IPs) for the same DP at a node
will be aggregated into a single IP \cite{yeh2014vip}. This is good
for efficiency, but also bad since we lose track of the actual expressed
demand once the suppression happens. To overcome this challenge, we
consider a \textit{dual-mode VIP framework}, which creates a virtual
plane (network) in which multiple interests are not suppressed via
the introduction of Virtual Interest Packets (VIPs), to keep a sufficient
statistic for the design of resouce control algorithm. As such, resource
control algorithms operating in the virtual plane can take advantage
of local popularity information on network demand (as represented
by the VIP counts), which is unavailable in the actual plane. Moreover,
this dual-mode-VIP-based approach also reduces the implementation
complexity of the VIP algorithm in the virtual plane considerably
(as compared with operating on DPs/IPs in the actual plane). 

\subsection{Transformation to a Virtual Network}

\begin{figure}
\begin{centering}
\includegraphics[width=85mm]{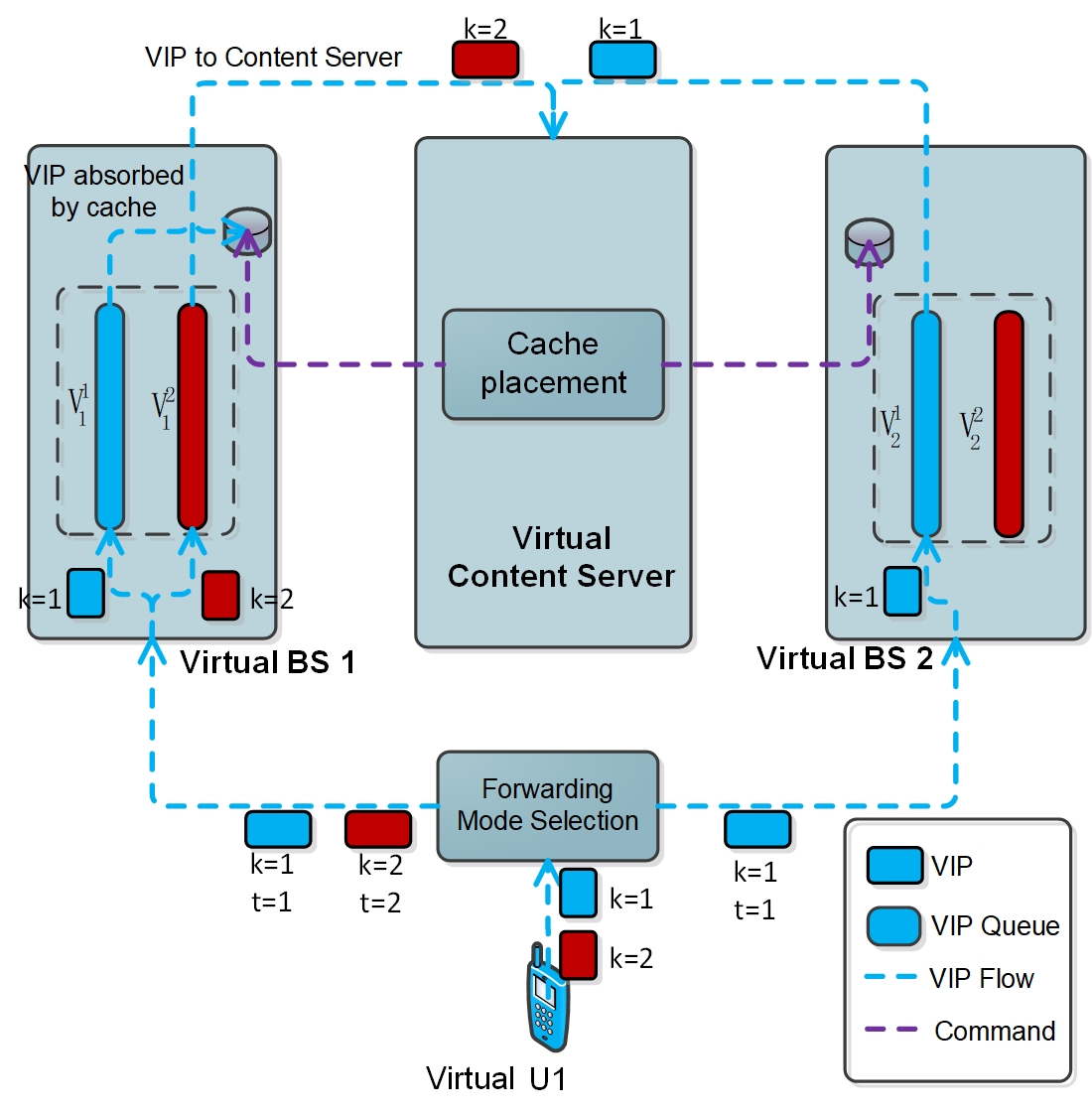}
\par\end{centering}
\caption{\label{fig:Vplane}{\small{}Illustration of the VIP flow in the virtual
network. In this example, user 1 requests data object 1 at time slot
1 and data object 2 at time slot 2. BS 1's cache contains data object
1 and BS 2's cache contains data object 2. The forwarding mode is
CoMP mode at time slot 1, and coordinated mode at time slot 2. Therefore,
the blue VIP with ID $k=1$ is forwarded to both BSs, and the red
VIP with ID $k=2$ is forwarded to BS 1 only. Moreover, BS 1 (2) forwards
the red (blue) VIP associated with data object 2 (1) to content server
since data object 2 (1) is not stored at BS1 (2).}}
\end{figure}

The dual-mode VIP framework relies on the concept of VIPs flowing
over a \textit{virtual network}, as illustrated in Fig. \ref{fig:Vplane}.
The virtual network is simulated at the content server and it has
exactly the same topology, cache state $\left\{ s_{n}^{k}(i)\right\} $
and global CSI $\boldsymbol{H}\left(t\right)$ as the actual network.
Each virtual node $m\in\mathcal{N}$ maintains a VIP queue $V_{m}^{k}(t)$
for each data object $k$, which is implemented as a counter in the
content server. The VIP queue $V_{m}^{k}(t)$ captures the local popularity
at each (virtual) node, and the set of all VIP queues $\mathbf{V}(t)=\left\{ V_{m}^{k}(t),\forall m\in\mathcal{M},k\in\mathcal{K}\right\} $
captures microscopic popularity variations. Initially, all VIP queues
are set to 0, i.e., $V_{m}^{k}(1)=0,\forall m\in\mathcal{M},k\in\mathcal{K}$.
As the content server receives data object requests (IPs requesting
the starting chunk of data objects) from users, the corresponding
VIP queues $V_{j}^{k}(t),j\in\mathcal{U}$ are incremented accordingly.
After some number of VIPs in $V_{j}^{k}(t),j\in\mathcal{U}$ have
been ``forwarded'' to the virtual BSs (in the virtual network),
the VIP queues $V_{j}^{k}(t),j\in\mathcal{U}$ are decreased and the
VIP queues $V_{n}^{k}(t),n\in\mathcal{B}$ are increased by the same
number accordingly. Similarly, after some number of VIPs in $V_{n}^{k}(t),n\in\mathcal{B}$
have been ``forwarded'' to the virtual content server (content source)
and local cache, the VIP queues $V_{n}^{k}(t),n\in\mathcal{B}$ are
decreased by the same number accordingly. Specifically, in the virtual
network, there are two modes for ``forwarding'' the VIPs from the
virtual users to virtual BSs, corresponding to the two PHY modes in
the actual plane. 

In the \textit{CoMP forwarding mode}, VIPs in $V_{j}^{k}(t)$ are
forwarded to all virtual BSs, and thus at time $t+1$, the VIP queues
become
\begin{align}
V_{j}^{k}(t+1) & =\left(V_{j}^{k}(t)-\mu_{j}^{Ak}(t)\right)^{+}+A_{j}^{k}(t),\nonumber \\
V_{n}^{k}(t+1) & =\left(\left(V_{n}^{k}(t)-\mu_{ng}^{k}(t)\right)^{+}+\sum_{j\in\mathcal{U}}\mu_{j}^{Ak}(t)-r_{n}s_{n}^{k}\left(i\right)\right)^{+}\label{eq:VIPCoMP}
\end{align}
$\forall j\in\mathcal{U},n\in\mathcal{B}$, where $A_{j}^{k}(t)$
is the number of exogenous data object request arrivals at the VIP
queue $V_{j}^{k}(t)$ during slot $t$, $\mu_{j}^{Ak}(t)$ is the
allocated transmission rate of VIPs for data object $k$ from virtual
user $j$ to all virtual BSs during time slot $t$ with CoMP forwarding
mode, $\mu_{ng}^{k}(t)$ is the allocated transmission rate of VIPs
for data object $k$ from virtual BS $n$ to the virtual content server
during time slot $t$, and $r_{n}$ is the maximum rate (in data objects/slot)
at which BS $n$ can produce copies of cached object $k$ (e.g., the
maximum rate $r_{n}$ may reflect the I/O rate of the storage disk).
On the other hand, in the coordinated forwarding mode, VIPs in $V_{j}^{k}(t)$
are ``forwarded'' to the serving virtual BS $n_{j}$ only, and thus
at time $t+1$, the VIP queues become
\begin{align}
V_{j}^{k}(t+1) & =\left(V_{j}^{k}(t)-\mu_{j}^{Bk}(t)\right)^{+}+A_{j}^{k}(t),\nonumber \\
V_{n}^{k}(t+1) & =\left(\left(V_{n}^{k}(t)-\mu_{ng}^{k}(t)\right)^{+}+\mu_{j_{n}}^{Bk}(t)-r_{n}s_{n}^{k}\left(i\right)\right)^{+},\label{eq:VIPcood}
\end{align}
$\forall j\in\mathcal{U},n\in\mathcal{B}$, where $\mu_{j}^{Bk}(t)$
is the allocated transmission rate of VIPs for data object $k$ from
virtual user $j$ to virtual BS $n_{j}$ during time slot $t$ with
coordinated forwarding mode. 

Combining the above two cases, the VIP queue dynamics can be expressed
in a compact form:
\begin{align}
V_{j}^{k}(t+1) & =\left(V_{j}^{k}(t)-\mu_{j}^{Ak}(t)M\left(t\right)-\mu_{j}^{Bk}(t)\overline{M}\left(t\right)\right)^{+}+A_{j}^{k}(t)\nonumber \\
V_{n}^{k}(t+1) & =\bigg(\left(V_{n}^{k}(t)-\mu_{ng}^{k}(t)\right)^{+}+\sum_{j\in\mathcal{U}}\mu_{j}^{Ak}(t)M\left(t\right)\nonumber \\
 & +\mu_{j_{n}}^{Bk}(t)\overline{M}\left(t\right)-r_{n}s_{n}^{k}\left(i\right)\bigg)^{+},\forall j\in\mathcal{U},n\in\mathcal{B},k\in\mathcal{K}\label{eq:VIPoverall}
\end{align}
where $M\left(t\right)\in\left\{ 0,1\right\} $ is the forwarding
mode at time slot $t$ in the virtual plane ($M\left(t\right)=0$
stands for the coordinated forwarding mode and $M\left(t\right)=1$
stands for the CoMP forwarding mode), and $\overline{M}\left(t\right)=\boldsymbol{1}_{M\left(t\right)=0}$.

In Table \ref{tab:notation}, we list the key notations in the actual
network and the corresponding notations in the virtual network for
easy reference.

\begin{table}
\begin{centering}
{\footnotesize{}}%
\begin{tabular}{|l|l|}
\hline 
{\small{}Notations in the actual network} & {\small{}Notations in the virtual network}\tabularnewline
\hline 
{\small{}PHY mode $M_{a}$ and IP modes $\left\{ m_{j}^{k}\right\} $} & {\small{}Forwarding mode $M$}\tabularnewline
{\small{}PHY rates $\left\{ \boldsymbol{c}^{A},\boldsymbol{c}^{B}\right\} $} & {\small{}Forwarding rates between BSs and users $\left\{ \boldsymbol{\mu}^{A},\boldsymbol{\mu}^{B}\right\} $}\tabularnewline
{\small{}Backhaul rates $\left\{ c_{ng}\right\} $} & {\small{}Forwarding rates between BSs and server $\left\{ \mu_{ng}^{k}\right\} $}\tabularnewline
{\small{}DP queues $\left\{ Q_{nj}^{A},Q_{nj_{n}}^{B},Q_{gn}\right\} $} & {\small{}VIP queues$\left\{ V_{n}^{k},V_{j}^{k}\right\} $}\tabularnewline
{\small{}IPs arrival rates $\left\{ a_{j}^{k}\right\} $} & {\small{}VIPs arrival rates $\left\{ A_{j}^{k}\right\} $}\tabularnewline
{\small{}Stability region $\Lambda_{c}$} & {\small{}VIP stability region $\Lambda_{v}$}\tabularnewline
\hline 
\end{tabular}{\footnotesize\par}
\par\end{centering}
{\small{}\caption{\label{tab:notation}{\small{}Key notations in the actual and virtual
networks.}}
}{\small\par}
\end{table}

\subsection{Mixed-Timescale Resource Control in Virtual Network\label{subsec:VIP-based-Resource-Control} }

A mixed-timescale resource control algorithm determines the slow-timescale
cache content placement policy $\left\{ p_{n}^{k}(i)\right\} $ and
the fast-timescale policies in the virtual network, aiming at solving
the following problem:
\begin{align}
\min_{\left\{ p_{n}^{k}(i),M\left(t\right),\mu_{j}^{Ak}(t),\mu_{j}^{Bk}(t),\mu_{ng}^{k}(t)\right\} }\limsup_{t\rightarrow\infty}\frac{1}{t}\sum_{\tau=1}^{t}\sum_{m\in\mathcal{M},k\in\mathcal{K}}\mathbb{E}\left[V_{m}^{k}(\tau)\right]+W\limsup_{J\rightarrow\infty}\frac{1}{J}\sum_{i=1}^{J}\mathbb{E}\left[\Gamma\left(i\right)\right],\nonumber \\
\text{s.t.}\sum_{k\in\mathcal{K}}s_{n}^{k}(i)\leq B_{C},\forall n\in\mathcal{B},\forall i,\nonumber \\
\sum_{k\in\mathcal{K}}\mu_{ng}^{k}(t)\leq R_{d},\forall n;\boldsymbol{\mu}^{A}\left(t\right)\in C^{A}\left(\boldsymbol{H}\right);\boldsymbol{\mu}^{B}\left(t\right)\in C^{B}\left(\boldsymbol{H}\right),\forall t,\label{eq:orgformu}
\end{align}
where the objective function is a weighted sum of the total average
VIP queue length and average cache placement cost, $W$ is a price
factor which can be used to control the tradeoff between the stability
(measured by the total average VIP queue length) and average cache
placement cost. However, it is difficult to directly solve this problem.
Therefore, we resort to the Lyapunov optimization framework \cite{neely2006energy,neely2010stochastic}
and approximately solve this problem by minimizing a mixed-timescale
drift-plus-penalty as follows.

\subsubsection{Slow-Timescale Cache Content Placement Solution}

Define $\mathcal{L}(\mathbf{V}(t))\triangleq\sum_{n\in\mathcal{N},k\in\mathcal{K}}\left(V_{n}^{k}(t)\right)^{2}$
as the Lyapunov function, which is a measure of unsatisfied requests
in the network. The cache content placement $\left\{ p_{n}^{k}(i)\right\} $
is designed to minimize the \textit{$T$-step VIP-drift-plus-penalty}
defined as
\begin{align}
\Delta_{T}\left(i\right)\triangleq\mathbb{E}\left[\mathcal{L}(\mathbf{V}(t_{0}^{i}+T))-\mathcal{L}(\mathbf{V}(t_{0}^{i}))|\boldsymbol{X}(t_{0}^{i})\right]\nonumber \\
+W\mathbb{E}\left[\sum_{n\in\mathcal{B},k\in\mathcal{K}}\gamma1_{\{p_{n}^{k}(i)=1\}}|\boldsymbol{X}(t_{0}^{i})\right],\label{eq:Tstepdrift}
\end{align}
where $t_{0}^{i}$ is the starting time slot of the $i$-th frame,
and $\boldsymbol{X}(t_{0}^{i})=\left[\mathbf{V}(t_{0}^{i})\,\boldsymbol{s}(i-1)\right]$
is the observed system state at the beginning of the frame. Intuitively,
if the first term in $\Delta_{T}\left(i\right)$ is negative, the
VIP lengths tend to decrease. On the other hand, the second term in
$\Delta_{T}\left(i\right)$ is the weighted cache content placement
cost, and $W$ is a price factor. Therefore, minimizing $\Delta_{T}\left(i\right)$
helps to strike a balance between stability and cost reduction. Following
a similar analysis to that in the proof of Lemma 3 in \cite{neely2005dynamic},
we obtain an upper bound of $\Delta_{T}\left(i\right)$. 
\begin{thm}
[$T$-step Drift-Plus-Penalty Upper Bound]\label{thm: T-step-Drift-Plus-Penalty-Upper}An
upper bound of $\Delta_{T}\left(i\right)$ is $\widetilde{\Delta}_{T}\left(i\right)\triangleq\mathbb{E}\left[\Delta_{T}^{U}\left(i\right)|\boldsymbol{X}(t_{0}^{i})\right]$,
where
\begin{align}
 & \Delta_{T}^{U}\left(i\right)=W\sum_{n\in\mathcal{B},k\in\mathcal{K}}\gamma1_{\{p_{n}^{k}(i)=1\}}+\overline{\Delta}\nonumber \\
 & -2T\sum_{n\in\mathcal{B},k\in\mathcal{K}}V_{n}^{k}(t_{0}^{i})r_{n}\left[s_{n}^{k}(i-1)+p_{n}^{k}(i)\right],\label{eq:Tdriftbound}
\end{align}
and $\overline{\Delta}$ is a term independent of $\left\{ p_{n}^{k}(i)\right\} $.

\end{thm}
\begin{IEEEproof}
The upper bound $\Delta_{T}^{U}\left(i\right)$ can be obstained by
manipulating the T-step virtual queue dynamics. For any $n\in\mathcal{B}$,
the virtual queue $V_{n}^{k}(t_{0}^{i}+T)$ at T step is bounded by,
\begin{align*}
 & V_{n}^{k}(t_{0}^{i}+T)\leq\bigg(V_{n}^{k}(t_{0}^{i})-\sum_{t=t_{0}^{i}}^{t_{0}^{i}+T-1}\mu_{ng}^{k}(t)-\sum_{t=t_{0}^{i}}^{t_{0}^{i}+T-1}r_{n}s_{n}^{k}\left(i\right)\bigg)^{+}\\
 & +\sum_{t=t_{0}^{i}}^{t_{0}^{i}+T-1}\bigg(\sum_{j\in\mathcal{U}}\mu_{j}^{Ak}(t)M\left(t\right)+\mu_{j_{n}}^{Bk}(t)\overline{M}\left(t\right)\bigg),\forall j\in\mathcal{U},n\in\mathcal{B},k\in\mathcal{K}
\end{align*}
By squaring both sides and applying the upper bounds on the transmission
rates, we have
\begin{align*}
 & \left(V_{n}^{k}(t_{0}^{i}+T)\right)^{2}-\left(V_{n}^{k}(t_{0}^{i})\right)^{2}\\
\leq & B-2TV_{n}^{k}(t_{0}^{i})r_{n}s_{n}^{k}(i)-2V_{n}^{k}(t_{0}^{i})\sum_{t=t_{0}^{i}}^{t_{0}^{i}+T-1}\mu_{ng}^{k}(t)\\
 & +2V_{n}^{k}(t_{0}^{i})\sum_{t=t_{0}^{i}}^{t_{0}^{i}+T-1}\bigg(\sum_{j\in\mathcal{U}}\mu_{j}^{Ak}(t)M\left(t\right)+\mu_{j_{n}}^{Bk}(t)\overline{M}\left(t\right)\bigg),
\end{align*}
where $B$ is a constant depending on $\left\{ \mu_{j,\text{max }}^{out},\mu_{n,\text{max }}^{out},\mu_{n,\text{max }}^{in}\right\} $
. We can obtain a similar bound for the drift of the VIP queue of
any user $j\in\mathcal{U}$. Summing over all the virtual queues for
all base stations and end users, we otain the T-step drift-plus penality
upper bound as in \ref{eq:Tdriftbound}, where $\overline{\Delta}$
is computed by aggregating the terms independent of $\left\{ p_{n}^{k}(i)\right\} $. 
\end{IEEEproof}
The slow-timescale drift minimization problem is given by \textit{
\begin{equation}
\min_{\left\{ p_{n}^{k}(i)\right\} }\Delta_{T}^{U}\left(i\right)\:\text{s.t. }\left(\ref{eq:cachesizecon}-\ref{eq:cacheover}\right)\text{ are satisfied}.\label{eq:Tstepmin}
\end{equation}
}The detailed steps to find the optimal solution of (\ref{eq:Tstepmin})
are summarized in Algorithm \ref{alg:Slow-timescale-cache}, which
only has linear complexity w.r.t. the number of data objects $K$.
In Algorithm \ref{alg:Slow-timescale-cache}, $V_{n}^{k}=V_{n}^{k}\left(t_{0}^{i}\right)$,
$\mathcal{C}$ is the set of currently cached data objects at BS $n$,\textbf{
$\mathcal{C}^{\prime}$} is a set of $B_{C}$ data objects with the
highest VIP counts (popularity), $\mathcal{O}^{\prime}=\mathcal{C^{\prime}}/\mathcal{C}$
is the set of the most popular data objects which have not been cached,
and $\mathcal{O}$ is the set of currently cached data objects which
are not in \textbf{$\mathcal{C}^{\prime}$}. Each data object $k_{i}^{'}$
in $\mathcal{O}^{\prime}$ will be added to the cache (i.e., $p_{n}^{k_{i}^{'}}(i)=1$)
if the benefit of caching it, as indicated by the backlog difference
$\left(V_{n}^{k_{i}^{'}}-V_{n}^{k_{i}}\right)r_{n}$, exceeds the
cache content placement cost threshold $\frac{W}{2T}\gamma$. If data
object $k_{i}^{'}$ is added to the cache and $i>\left(\left|\mathcal{O}^{\prime}\right|-\left|\mathcal{O}\right|\right)^{+}$,
data object $k_{i}$ in $\mathcal{O}$ will be removed (i.e., $p_{n}^{k_{i}}(i)=-1$)
to save space for caching data object $k_{i}^{'}$. 
\begin{prop}
Algorithm \ref{alg:Slow-timescale-cache} finds the optimal solution
of Problem (\ref{eq:Tstepmin}). 
\end{prop}
\begin{IEEEproof}
The proof can be obtained by contradiction. Suppose a certain content
$k_{i}$ is not chosen to be cached in Algorithm \ref{alg:Slow-timescale-cache},
and by caching content $k_{i}^{'}$ (i.e., $p_{n}^{k_{i}^{'}}(i)=1$)
and removing content $k_{j}$ (i.e., $p_{n}^{k_{j}}(i)=-1$), $\Delta_{T}^{U}\left(i\right)$
can be further lowered. Then the backlog reduction of the content
is $\left(V_{n}^{k_{i}^{'}}-V_{n}^{k_{j}}\right)r_{n}$, which should
be larger than that of any cached content chosen by Algorithm \ref{alg:Slow-timescale-cache}
and the threshold $\frac{W}{2T}\gamma$, to decrease the total drift.
However, Algorithm \ref{alg:Slow-timescale-cache} works by caching
the contents with the largest backlog differences $\left(V_{n}^{k_{i}^{'}}-V_{n}^{k_{j}}\right)r_{n}$
as guaranted by the sorting operations, and the content not cached
by Algorithm \ref{alg:Slow-timescale-cache} cannot have a larger
backlog difference, which yields the contradiction. 
\end{IEEEproof}
\begin{algorithm}
\caption{\label{alg:Slow-timescale-cache}Slow-timescale cache content placement
solution at frame $i$}

{\small{}For each base station $n$, }{\small\par}
\begin{itemize}
\item {\small{}Let $\mathcal{C}=\left\{ k|s_{n}^{k}(i-1)=1,k\in\mathcal{K}\right\} $.}{\small\par}
\item {\small{}Let $\mathcal{C^{\prime}}=\left\{ k|s_{n}^{k\star}(i)=1\right\} $,
where $\left\{ s_{n}^{k\star}(i)\right\} $ is the optimal solution
of $\max_{\left\{ s_{n}^{k}\right\} }\sum_{k\in\mathcal{K}}V_{n}^{k}s_{n}^{k},\,\sum_{k\in\mathcal{K}}s_{n}^{k}\leq B_{C}$.}{\small\par}
\item {\small{}Let $\mathcal{O}^{\prime}=\mathcal{C^{\prime}}/\mathcal{C}$,
$\mathcal{O=C}/\mathcal{C^{\prime}}$.}\\
{\small{}Sort the queue backlogs as, $V_{n}^{k_{1}^{'}}\geq\cdots\geq V_{n}^{k_{\left|\mathcal{O}^{\prime}\right|}^{'}}$,
$0\leq0\leq\cdots\leq V_{n}^{k_{\left|\mathcal{O}^{\prime}\right|-|\mathcal{O}|+1}}\leq\cdots\leq V_{n}^{k_{\left|\mathcal{O}^{\prime}\right|}}$.}{\small\par}
\item {\small{}For $i=1:\left(\left|\mathcal{O}^{\prime}\right|-\left|\mathcal{O}\right|\right)^{+}$,
if $V_{n}^{k_{i}^{'}}r_{n}\geq\frac{W}{2T}\gamma$, then let $p_{n}^{k_{i}^{'}}(i)=1$. }{\small\par}
\item {\small{}For $i=\left(\left|\mathcal{O}^{\prime}\right|-\left|\mathcal{O}\right|\right)^{+}+1:\left|\mathcal{O}^{\prime}\right|$,
if $\left(V_{n}^{k_{i}^{'}}-V_{n}^{k_{i}}\right)r_{n}\geq\frac{W}{2T}\gamma$,
then let $p_{n}^{k_{i}^{'}}(i)=1$ and $p_{n}^{k_{i}}(i)=-1$. }{\small\par}
\end{itemize}
\end{algorithm}

\subsubsection{Fast-Timescale Control Solution }

Similarly, the fast-timescale control solution $\left\{ M\left(t\right)\right\} $
and $\left\{ \mu_{j}^{Ak}(t),\mu_{j}^{Bk}(t),\mu_{ng}^{k}(t)\right\} $
is obtained by solving the following \textit{$1$-step VIP-drift minimization
problem:}
\begin{align}
\min_{\begin{array}{c}
\left\{ M\left(t\right),\mu_{ng}^{k}(t)\right\} \\
\left\{ \mu_{j}^{Ak}(t),\mu_{j}^{Bk}(t)\right\} 
\end{array}}\:M\left(t\right)\sum_{j\in\mathcal{U},k\in\mathcal{K}}\mu_{j}^{Ak}(t)\left(\sum_{n\in\mathcal{B}}V_{n}^{k}(t)-V_{j}^{k}(t)\right)\label{eq:1stepmin}\\
+\overline{M}\left(t\right)\sum_{j\in\mathcal{U},k\in\mathcal{K}}\mu_{j}^{Bk}(t)\left(V_{n_{j}}^{k}(t)-V_{j}^{k}(t)\right)-\sum_{n\in\mathcal{B},k\in\mathcal{K}}V_{n}^{k}(t)\mu_{ng}^{k}(t)\nonumber \\
\text{s.t.}\sum_{k\in\mathcal{K}}\mu_{ng}^{k}(t)\leq R_{d},\forall n;\boldsymbol{\mu}^{A}\left(t\right)\in C^{A}\left(\boldsymbol{H}\right);\boldsymbol{\mu}^{B}\left(t\right)\in C^{B}\left(\boldsymbol{H}\right)\label{eq:linkcap}
\end{align}
where $\boldsymbol{\mu}^{A}\left(t\right)=\left[\sum_{k\in\mathcal{K}}\mu_{j}^{Ak}\left(t\right)\right]_{j\in\mathcal{U}}\in\mathbb{R}_{+}^{N}$,
$\boldsymbol{\mu}^{B}\left(t\right)=\left[\sum_{k\in\mathcal{K}}\mu_{j}^{Bk}\left(t\right)\right]_{j\in\mathcal{U}}\in\mathbb{R}_{+}^{N}$.
Note that (\ref{eq:linkcap}) is the link capacity constraint in the
virtual plane. The detailed steps to find the optimal solution of
(\ref{eq:1stepmin}) are summarized in Algorithm \ref{alg:Fast-timescacle-control}.
In step 1 and 2, we need to solve two weighted sum-rate maximization
problems in a MIMO BC and IFC, respectively. There are many existing
weighted sum-rate maximization algorithms for different CoMP/coordinated
transmission schemes \cite{Luo_TSP11_WMMSE} but the details are omitted
for conciseness.

\begin{algorithm}
\caption{\label{alg:Fast-timescacle-control}Fast-timescale control solution
at slot $t$}

\textbf{\small{}1. Backhaul rate allocation}{\small\par}

\textbf{\small{}Let}{\small{} $\mu_{ng}^{k}(t)=\begin{cases}
R_{d} & k=k_{n}^{*}(t)\\
0 & otherwise
\end{cases},\forall n\in\mathcal{B}$, where $k_{n}^{*}(t)\triangleq\arg\max_{k}V_{n}^{k}(t)$.}{\small\par}

\textbf{\small{}2. Forwarding mode selection and rate allocation}{\small\par}

\textbf{\small{}Let}{\small{} 
\begin{align}
\left\{ \mu_{j}^{Ak*}(t)\right\}  & =\underset{\left\{ \mu_{j}^{Ak}(t)\right\} }{\text{argmax}}\sum_{j\in\mathcal{U},k\in\mathcal{K}}\mu_{j}^{Ak}(t)\left(V_{j}^{k}(t)-\sum_{n\in\mathcal{B}}V_{n}^{k}(t)\right)\label{eq:WSRA}\\
 & \text{s.t. }\boldsymbol{\mu}^{A}\left(t\right)\in C^{A}\left(\boldsymbol{H}\right)\nonumber 
\end{align}
\begin{align}
\left\{ \mu_{j}^{Bk*}(t)\right\}  & =\underset{\left\{ \mu_{j}^{Bk}(t)\right\} }{\text{argmax}}\sum_{j\in\mathcal{U},k\in\mathcal{K}}\mu_{j}^{Bk}(t)\left(V_{j}^{k}(t)-V_{n_{j}}^{k}(t)\right)\label{eq:WSRB}\\
 & \text{s.t. }\boldsymbol{\mu}^{B}\left(t\right)\in C^{B}\left(\boldsymbol{H}\right).\nonumber 
\end{align}
}{\small\par}

\textbf{\small{}Let}{\small{} $\Delta_{1}^{A}=\sum_{j\in\mathcal{U},k\in\mathcal{K}}\mu_{j}^{Ak*}(t)\left(V_{j}^{k}(t)-\sum_{n\in\mathcal{B}}V_{n}^{k}(t)\right)$
and $\Delta_{1}^{B}=\sum_{j\in\mathcal{U},k\in\mathcal{K}}\mu_{j}^{Bk*}(t)\left(V_{j}^{k}(t)-V_{n_{j}}^{k}(t)\right)$.}{\small\par}

\textbf{\small{}If}{\small{} $\Delta_{1}^{A}\geq\Delta_{1}^{B}$,
}\textbf{\small{}let}{\small{} $M\left(t\right)=1$, $\left\{ \mu_{j}^{Ak}(t)\right\} =\left\{ \mu_{j}^{Ak*}(t)\right\} $,
$\mu_{j}^{Bk}(t)=0,\forall j,k$; }{\small\par}

\textbf{\small{}Else}{\small{}, }\textbf{\small{}let}{\small{} $M\left(t\right)=0$,
$\mu_{j}^{Ak}(t)=0,\forall j,k$, $\left\{ \mu_{j}^{Bk}(t)\right\} =\left\{ \mu_{j}^{Bk*}(t)\right\} $.}{\small\par}
\end{algorithm}

\subsection{Virtual-to-Actual Control Policy Mapping\label{subsec:Virtual-to-Actual-Control-Policy}}

In the following, we propose a \textit{virtual-to-actual control policy
mapping} which can generate a resource control policy for the actual
network from that in the virtual network.\textbf{ }

\textbf{Mapping for cache placement control policy $\left\{ p_{n}^{k}(i)\right\} $:}
The cache placement control action in the actual network is the same
as that in the virtual network.

\textbf{Mapping for IP mode selection policy }$\left\{ m_{j}^{k}\left(t\right)\right\} $\textbf{:}
For a given forwarding mode selection and rate allocation policy $\left\{ M\left(t\right),\mu_{j}^{Ak}(t)\right\} $
in the virtual network that achieves an average transmission rate
of VIPs $\overline{\mu}_{j}^{Ak}=\limsup_{t\rightarrow\infty}\frac{1}{t}\sum_{\tau=1}^{t}M\left(\tau\right)\mu_{j}^{Ak}(\tau)$
for data object $k$ from virtual user $j$ to all virtual BSs, we
need to construct an IP mode selection policy $\left\{ m_{j}^{k}\left(t\right)\right\} $
in the actual network such that the same average transmission rate
of IPs for data object $k$ from user $j$ to all BSs can be achieved,
i.e., 
\begin{equation}
\limsup_{t\rightarrow\infty}\frac{1}{t}\sum_{\tau=1}^{t}m_{j}^{k}\left(\tau\right)a_{j}^{k}(\tau)=\overline{\mu}_{j}^{Ak}.\label{eq:vaflowblan}
\end{equation}
To achieve this, the content server maintains a set of \textit{virtual
CoMP queues} as
\begin{equation}
U_{j}^{Ak}(t+1)=U_{j}^{Ak}(t)-m_{j}^{k}\left(t\right)a_{j}^{k}(t)+M\left(t\right)\mu_{j}^{Ak}(t),\forall j,k,\label{eq:VComPQ}
\end{equation}
where $U_{j}^{Ak}(1)=0,\forall j,k$. Clearly, by setting the forwarding
mode in the actual plane as $m_{j}^{k}\left(t\right)=\boldsymbol{1}_{U_{j}^{Ak}(t+1)>0},\forall t$,
we can achieve a bounded $\left|U_{j}^{Ak}(t+1)\right|,\forall t$
(since $\mu_{j}^{Ak}(t)$ is bounded), which implies that (\ref{eq:vaflowblan})
can also be satisfied. 

\textbf{Mapping for PHY mode selection and rate allocation policy:}
For the PHY mode selection and rate allocation policy in the actual
plane, we let $c_{ng}(t)=\sum_{k\in\mathcal{K}}\mu_{ng}^{k}(t)$,
$M_{a}\left(t\right)=M\left(t\right)$, $\boldsymbol{c}^{A}\left(t\right)=\boldsymbol{\mu}^{A}\left(t\right)$
and $\boldsymbol{c}^{B}\left(t\right)=\boldsymbol{\mu}^{B}\left(t\right)$. 
\begin{rem}
The proposed dual-mode VIP framework has three important differences
from the existing VIP framework in \cite{yeh2014vip}. First, there
are two different forwarding modes in the virtual network. Second,
all VIP dynamics are maintained centrally at the content server. This
is necessary for wireless networks where centralized resource control
(within a CoMP cluster) is required to mitigate the interference and
fading channel effects. Third, all actions, such as ``forward''
or ``send'', in the virtual network are merely some calculations
performed at the content server to ``simulate'' the VIP flows and
dynamics. They do not generate signaling overhead in the actual network. 
\end{rem}

\section{Throughput Optimality Analysis\label{sec:Throughput-Optimality-Analysis}}

In this section, we first introduce the concept of the \textit{network
stability region} \textit{under conditional flow balance}. Then we
establish the equivalence between the virtual and actual networks,
and the throughput optimality of the proposed VIP-based resource control
algorithm. For all the theoretical analysis in this and the next section,
we make the following standard assumptions on the arrival processes
$A_{j}^{k}(t)$'s: (i) The arrival processes $\left\{ A_{j}^{k}(t);t=1,2,...\right\} $
are mutually independent with respect to $j$ and $k$; and (ii) for
all $j\in\mathcal{U},k\in\mathcal{K}$, $\left\{ A_{j}^{k}(t);t=1,2,...\right\} $
are i.i.d. with respect to $t$ and $A_{j}^{k}(t)\leq A_{max}^{k}$
for all $t$. 

\subsection{Motivation of Conditional Flow Balance}

In practice, a basic QoS requirement is to maintain the queue stability
for the data flow of each user. Specifically, a queue $Q\left(t\right)$
is stable if 
\begin{equation}
\overline{Q}\triangleq\limsup_{t\rightarrow\infty}\frac{1}{t}\sum_{\tau=1}^{t}\mathbb{E}\left[Q\left(\tau\right)\right]<\infty,\label{eq:Qbar}
\end{equation}
where $\overline{Q}$ is called the \textit{limiting average queue
length} of $Q\left(t\right)$. A necessary condition for a $Q\left(t\right)$
with dynamics $Q\left(t+1\right)=\left(Q\left(t\right)-b(t)\right)^{+}+a(t)$
to be stable is that the following \textit{flow balance constraint}
is satisfied:
\begin{equation}
\limsup_{t\rightarrow\infty}\frac{1}{t}\sum_{\tau=1}^{t}\mathbb{E}\left[a\left(\tau\right)-b(\tau)\right]\leq0.\label{eq:flowbal}
\end{equation}
The conventional \textit{network stability region} $\Lambda$ is defined
as the closure of the set of all arrival rate tuples $\boldsymbol{\lambda}=\left(\lambda_{j}^{k}\right)_{j\in\mathcal{U},k\in\mathcal{K}}$
for which there exists some resource control policy which can guarantee
that all data queues are stable, where $\lambda_{j}^{k}=\lim_{t\rightarrow\infty}\frac{1}{t}\sum_{\tau=1}^{t}A_{j}^{k}(\tau)$
is the long-term exogenous VIP arrival rate at the VIP queue $V_{j}^{k}(t)$.
To guarantee the basic QoS requirements of all users, the system should
not operate at a point outside the stability region. 

However, the flow balance or queue stability constraint is not sufficient
to guarantee a good performance for practical cached interference
networks, as explained below. In cached networks, the cache state
$\boldsymbol{s}\left(i\right)$ has a huge impact on the arrival rate
of $Q_{gn}\left(t\right)$. Specifically, let $\mathcal{K}_{n}$ denote
the subset of data objects that need to be delivered to BS $n$. When
$s_{n}^{k}\left(i\right)=1,\forall k\in\mathcal{K}_{n}$, all the
requests about data objects in $\mathcal{K}_{n}$ will be served by
the local cache at BS $n$, and thus $b_{n}\left(t\right)=0$. When
$s_{n}^{k}\left(i\right)=0,\forall k\notin\mathcal{K}_{n}$, all the
requests about data objects in $\mathcal{K}_{n}$ will be forwarded
to the content server, and thus $b_{n}\left(t\right)$ is large. Recall
that in practice, $\boldsymbol{s}\left(i\right)$ is a \textit{slow-timescale
process} which can only change at the timescale of frames ($T\gg1$
time slots) to avoid frequent cache content placement. If only the
flow balance or queue stability constraint is considered, $s_{n}^{k}(i),\forall k\in\mathcal{K}_{n}$
may remain $0$ for several frames, during which $Q_{gn}\left(t\right)$
may keep growing. As a result, the average delay of DPs would be in
the order of several frames, which is unacceptable in practice. To
address this issue, we introduce the concept of the \textit{network
stability region under conditional flow balance} $\Lambda_{c}\subset\Lambda$,
as will be formally defined in the next subsection. When the arrival
rates $\left(\lambda_{j}^{k}\right)\in\Lambda_{c}$, there exists
some resource control policy which can guarantee that the flow balance
is satisfied conditioned on any \textit{cache state} $\boldsymbol{s}$
of non-zero probability. This stronger notion of stability ensures
that the system will not operate at a point with excessively large
delay. Besides the above practical consideration, imposing the conditional
flow balance constraint also makes it more tractable to establish
the throughput optimaility of the proposed algorithm. 

\subsection{Stability Region under Conditional Flow Balance}

The \textit{limiting probability} that a cache state $\boldsymbol{s}$
occurs is
\begin{equation}
\pi_{\boldsymbol{s}}=\limsup_{i\rightarrow\infty}\mathbb{E}\left[\frac{\left|\mathcal{T}\left(\boldsymbol{s},i\right)\right|}{i}\right],\label{eq:StedProb}
\end{equation}
where $\mathcal{T}\left(\boldsymbol{\boldsymbol{s}},i\right)=\left\{ \tau\leq i:\:\boldsymbol{s}\left(\tau\right)=\boldsymbol{s}\right\} $.
Let $\mathcal{S}=\left\{ \boldsymbol{s}:\:\pi_{\boldsymbol{s}}>0\right\} $
denote the set of all cache states with non-zero limiting probability.
Then the \textit{conditional flow balance constraint} is
\begin{equation}
\overline{b}_{n|\boldsymbol{s}}\leq\overline{c}_{ng|\boldsymbol{s}},\overline{b}_{nj_{n}|\boldsymbol{s}}^{B}\leq\overline{c}_{j_{n}|\boldsymbol{s}}^{B},\overline{b}_{nj|\boldsymbol{s}}^{A}\leq\overline{c}_{j|\boldsymbol{s}}^{A}\label{eq:condfloworg}
\end{equation}
$\forall j\in\mathcal{U},n\in\mathcal{B}$ and $\forall\boldsymbol{s}\in\mathcal{S}$,
where $\left(\overline{b}_{n|\boldsymbol{s}},\overline{c}_{ng|\boldsymbol{s}}\right)$,
$\left(\overline{b}_{nj_{n}|\boldsymbol{s}}^{B},\overline{c}_{j_{n}|\boldsymbol{s}}^{B}\right)$
and $\left(\overline{b}_{nj|\boldsymbol{s}}^{A}\overline{c}_{j|\boldsymbol{s}}^{A}\right)$
are $\boldsymbol{s}$\textit{-conditional average rates} of $\left(b_{n}\left(t\right),c_{ng}(t)\right)$,
$\left(b_{nj_{n}}^{B}(t),\overline{M}_{a}\left(t\right)c_{j_{n}}^{B}\left(t\right)\right)$
and $\left(b_{nj}^{A}(t),M_{a}\left(t\right)c_{j}^{A}\left(t\right)\right)$
(arrival/departure rates of $Q_{gn}\left(t\right)$, $Q_{nj_{n}}^{B}\left(t\right)$
and $Q_{nj}^{A}\left(t\right)$, respectively), i.e., average rates
over all frames with cache state $\boldsymbol{s}$. For example, the
$\boldsymbol{s}$\textit{-conditional average departure rate} of $Q_{nj_{n}}^{B}\left(t\right)$
is defined as
\begin{equation}
\overline{c}_{j_{n}|\boldsymbol{s}}^{B}=\limsup_{i\rightarrow\infty}\mathbb{E}\left[\frac{\sum_{\tau\in\mathcal{T}\left(\boldsymbol{\boldsymbol{s}},i\right)}}{T\left|\mathcal{T}\left(\boldsymbol{\boldsymbol{s}},i\right)\right|}\sum_{t=1+(\tau-1)T}^{\tau T}\overline{M}_{a}\left(t\right)c_{j_{n}}^{B}\left(t\right)\right].\label{eq:scondrate}
\end{equation}
The other $\boldsymbol{s}$\textit{-conditional average rates }are
defined similarly.
\begin{defn}
The\textit{ network stability region} \textit{under conditional flow
balance} $\Lambda_{c}$ is the closure of the set of all arrival rate
tuples $\boldsymbol{\lambda}$ for which there exists some resource
control policy which can guarantee that all data queues are stable
and also satisfies the cache size constraint (\ref{eq:cachesizecon}),
conditional flow balance constraint (\ref{eq:condfloworg}), and the
following link capacity constraint:
\begin{equation}
c_{ng}(t)\leq R_{d},\forall n;\boldsymbol{c}^{A}\left(t\right)\in C^{A}\left(\boldsymbol{H}\right);\boldsymbol{c}^{B}\left(t\right)\in C^{B}\left(\boldsymbol{H}\right);\forall t.\label{eq:linkCap}
\end{equation}
\end{defn}

Similarly, we can define the \textit{VIP} \textit{stability region}
\textit{under conditional flow balance}. In the virtual plane, the
conditional flow balance constraint is
\begin{equation}
\lambda_{j}^{k}\leq\overline{\mu}_{j|\boldsymbol{s}}^{Ak}+\overline{\mu}_{j|\boldsymbol{s}}^{Bk};\sum_{j\in\mathcal{U}}\overline{\mu}_{j|\boldsymbol{s}}^{Ak}+\overline{\mu}_{j_{n}|\boldsymbol{s}}^{Bk}\leq\overline{\mu}_{ng|\boldsymbol{s}}^{k}+r_{n}s_{n}^{k},\label{eq:condfloworgV}
\end{equation}
$\forall j\in\mathcal{U},k\in\mathcal{K},n\in\mathcal{B}$ and $\forall\boldsymbol{s}\in\mathcal{S}$,
where $\overline{\mu}_{j|\boldsymbol{s}}^{Ak},\overline{\mu}_{j|\boldsymbol{s}}^{Bk}$
and $\overline{\mu}_{ng|\boldsymbol{s}}^{k}$ are $\boldsymbol{s}$\textit{-conditional
average rates }of $\mu_{j}^{Ak}(t)M\left(t\right)$, $\mu_{j}^{Bk}(t)\overline{M}\left(t\right)$
and $\mu_{ng}^{k}(t)$, whose definitions are similar to (\ref{eq:scondrate}).
\begin{defn}
The\textit{ VIP stability region} \textit{under conditional flow balance}
$\Lambda_{v}$ is the closure of the set of all arrival rate tuples
$\boldsymbol{\lambda}$ for which there exists some virtual resource
control policy which makes all VIP queues stable and satisfies the
cache size constraint (\ref{eq:cachesizecon}), conditional flow balance
constraint (\ref{eq:condfloworgV}), and link capacity constraint
(\ref{eq:linkcap}) in the virtual network.
\end{defn}

In the rest of the paper, ``the stability region'' always refers
to the stability region under conditional flow balance.

\subsection{Equivalence between the Virtual and Actual Networks\label{subsec:Equivalence-between-VA}}

Unlike the virtual network, where each data object $k$ corresponds
to an individual VIP queue, each DP queue in the actual network contains
DPs of all data objects. Hence, the virtual network is not exactly
a ``time reversal mirror'' (TRM) of the actual network, and it is
non-trivial to establish the equivalence between them. This challenge
is addressed in the following theorem, which is proved in Appendix
\ref{subsec:Proof-of-TheoremVAE}.
\begin{thm}
[Equivalence between virtual and actual networks]\label{thm:Equivalence-between-VA}$\Lambda_{c}=\Lambda_{v}$,
and for any arrival rate tuple $\boldsymbol{\lambda}\in\textrm{int}\Lambda_{v}$,
we have $\overline{\Gamma}_{c}^{*}\left(\boldsymbol{\lambda}\right)=\overline{\Gamma}_{v}^{*}\left(\boldsymbol{\lambda}\right)$,
where $\overline{\Gamma}_{c}^{*}\left(\boldsymbol{\lambda}\right)$
and $\overline{\Gamma}_{v}^{*}\left(\boldsymbol{\lambda}\right)$
are the minimum cache content placement costs required for stability
in the actual and virtual networks respectively, under the arrival
rate tuple $\boldsymbol{\lambda}$. 
\end{thm}

\subsection{Optimality of the Dual-mode-VIP-based Resource Control}

The throughput optimality of the proposed resource control is summarized
below.
\begin{thm}
[Throughput optimality for the virtual network]\label{thm:Throughput-optimality-forVN}If
there exists $\bm{\epsilon}=\left(\epsilon_{n}^{k}=\epsilon\right){}_{n\in\mathcal{N},k\in\mathcal{K}}\succ\mathbf{0}$
such that $\bm{\lambda+\epsilon}\in\Lambda_{v}$, then the VIP queues
and cache content placement cost under the mixed-timescale resource
control algorithm in Section \ref{subsec:VIP-based-Resource-Control}
(Algorithms \ref{alg:Slow-timescale-cache} and \ref{alg:Fast-timescacle-control})
satisfies
\begin{align}
\limsup_{t\rightarrow\infty} & \frac{1}{t}\sum_{\tau=1}^{t}\sum_{m\in\mathcal{M},k\in\mathcal{K}}\mathbb{E}\left[V_{m}^{k}(\tau)\right]\leq\frac{TB+W\overline{\Gamma}_{v}^{*}\left(\boldsymbol{\lambda}\right)}{\epsilon},\nonumber \\
\limsup_{J\rightarrow\infty} & \frac{1}{J}\sum_{i=1}^{J}\mathbb{E}\left[\Gamma\left(i\right)\right]\leq\frac{TB}{W}+\overline{\Gamma}_{v}^{*}\left(\boldsymbol{\lambda}\right),\label{eq:Thpopt}
\end{align}
where $B$ is a constant depending on the maximum endogenous rates
$\mu_{j,\text{max }}^{out}\triangleq\underset{\boldsymbol{H},\boldsymbol{\mu}^{A}}{\max}\mu_{j}^{A},\text{ s.t. }\boldsymbol{\mu}^{A}\in C^{A}\left(\boldsymbol{H}\right)$,$\forall j\in\mathcal{U}$,
$\mu_{n,\text{max }}^{out}\triangleq R_{d},\forall n\in\mathcal{B}$,
maximum exogenous rates $\mu_{n,\text{max }}^{in}\triangleq\underset{\boldsymbol{H},\boldsymbol{\mu}^{A}}{\max}\sum_{j\in\mathcal{U}}\mu_{j}^{A},\text{ s.t. }\boldsymbol{\mu}^{A}\in C^{A}\left(\boldsymbol{H}\right)$
and the maximum arrival rate $A_{\text{max}}$ at each user. 
\end{thm}

Please refer to Appendix \ref{subsec:Proof-of-TheoremThpopt} for
the proof. Theorem \ref{thm:Throughput-optimality-forVN} states that
the proposed solution is throughput-optimal for the virtual network
since for any $\boldsymbol{\lambda}\in\textrm{int}\Lambda_{v}$, the
average cost can be made arbitrarily close to the minimum cost $\overline{\Gamma}_{v}^{*}\left(\boldsymbol{\lambda}\right)$
with bounded VIP queue lengths by choosing a sufficiently large $W$.
Note that the virtual-to-actual control policy mapping in Section
\ref{subsec:Virtual-to-Actual-Control-Policy} is designed to satisfy
the following property: if the virtual resource control policy satisfies
the conditional flow balance constraint (\ref{eq:condfloworgV}),
the resulting resource control policy also satisfies the conditional
flow balance constraint (\ref{eq:condfloworg}) in the actual network.
Therefore, Theorem \ref{thm:Throughput-optimality-forVN} implies
that the proposed solution is also throughput-optimal for the actual
network.

\section{Characterization of the Stability Region\label{sec:Characterization-of-the}}

Since the network stability region is equal to the VIP stability region,
we shall focus on the characterization of the VIP stability region,
which is easier.

\subsection{VIP DoF Stability Region under Identical User Preference}

The VIP stability region for the general case is given in Lemma \ref{lem:VIP-stability-region}
in Appendix \ref{subsec:VIP-Stability-Region}. To obtain insight
for practical design, we focus on studying the \textit{VIP DoF stability
region under identical user preference} defined as
\begin{equation}
\mathcal{D}_{ve}\triangleq\lim_{P\rightarrow\infty}\left(\Lambda_{v}\cap\Lambda_{e}\right)/P,\label{eq:VIPDOF}
\end{equation}
where $\Lambda_{e}=\left\{ \boldsymbol{\lambda}:\:\lambda_{j}^{1}\geq\lambda_{j}^{2}\geq...\geq\lambda_{j}^{K},\forall j\right\} $.
$\mathcal{D}_{ve}$ captures the VIP stability region when the SNR
is high and the popularity orders of the $K$ data objects at all
users are identical.
\begin{thm}
[VIP DoF stability region]\label{thm:VIP-stability-region-closed}For
any arrival rate tuple $\boldsymbol{\lambda}\in\textrm{int}\left(\Lambda_{v}\cap\Lambda_{e}\right)$,
the minimum cache content placement cost required for stability is
given by $\overline{\Gamma}_{v}^{*}\left(\boldsymbol{\lambda}\right)=0$,
which is achieved by a fixed cache placement $s_{n}^{k}=1,\forall k\in\mathcal{K}_{p},n$,
and $s_{n}^{k}=0,\forall k\in\overline{\mathcal{K}}_{p},n$, where
$\mathcal{K}_{p}=\left\{ 1,...,L_{C}\right\} $ and $\overline{\mathcal{K}}_{p}=\mathcal{K}\backslash\mathcal{K}_{p}$.
Moreover, if 
\begin{equation}
\lim_{P\rightarrow\infty}C^{A}\left(\boldsymbol{H}\right)/P=\mathcal{D}^{A},\lim_{P\rightarrow\infty}C^{B}\left(\boldsymbol{H}\right)/P=\mathcal{D}^{B},\forall\boldsymbol{H},\label{eq:DoFaSM}
\end{equation}
where $\mathcal{D}^{A}$ and $\mathcal{D}^{B}$ are DoF regions under
the CoMP and coordinated modes respectively, then $\mathcal{D}_{ve}$
consists of all DoF tuples $\boldsymbol{d}=\left(d_{j}^{k}\right)_{k\in\mathcal{K},j\in\mathcal{U}}$
such that there exists a set of variables $\left\{ \alpha\in\left[0,1\right],\mu_{j}^{Ak}\geq0,\mu_{j}^{Bk}\geq0,\mu_{ng}^{k}\geq0\right\} $
satisfying
\begin{align}
 & d_{j}^{k}\leq\left(1-\alpha\right)\mu_{j}^{Ak}+\alpha\mu_{j}^{Bk},\forall j,k\nonumber \\
 & \left(\sum_{k\in\mathcal{K}_{p}}\left[\left(1-\alpha\right)\sum_{j\in\mathcal{U}}\mu_{j}^{Ak}+\alpha\mu_{j_{n}}^{Bk}\right]-r_{n}\right)^{+}\nonumber \\
 & +\sum_{k\in\overline{\mathcal{K}}_{p}}\left[\left(1-\alpha\right)\sum_{j\in\mathcal{U}}\mu_{j}^{Ak}+\alpha\mu_{j_{n}}^{Bk}\right]\leq R_{d},\forall n\nonumber \\
 & \left[\sum_{k\in\mathcal{K}}\mu_{j}^{Ak}\right]_{j\in\mathcal{U}}\in\mathcal{D}^{A},\left[\sum_{k\in\mathcal{K}}\mu_{j}^{Bk}\right]_{j\in\mathcal{U}}\in\mathcal{D}^{B}.\label{eq:djk}
\end{align}
\end{thm}

Please refer to Appendix \ref{subsec:VIP-Stability-Region} for the
proof. Note that (\ref{eq:DoFaSM}) is a mild conditon because it
holds for many channel distributions (such as Rayleigh or Rice fading
channels). Moreover, the DoF region is determined by the distribution
of $\boldsymbol{H}$ instead of the realization of $\boldsymbol{H}$
\cite{Poor_TIT2013_MIMOIFCDOF}. Therefore, $\mathcal{D}^{A}$ and
$\mathcal{D}^{B}$ are not expresssed as a function of $\boldsymbol{H}$.
 
\begin{rem}
For the special case in Theorem \ref{thm:VIP-stability-region-closed}
with stationary popularity and identical user preference, the fixed
offline cache placement (caching the most popular $L_{C}$ data objects)
is sufficient to achieve the minimum cache content placement cost
$\overline{\Gamma}_{v}^{*}\left(\boldsymbol{\lambda}\right)=0$. This
is because, in this special case, each user at each frame sees the
same stationary arrival rate process $\left\{ A_{j}^{k}(t)\right\} $.
In order to satsify the conditional flow balance constraint within
each frame, we should cache the most popular $L_{C}$ data objects
to induce CoMP to handle the large arrival rates caused by the more
frequent requests of popular data objects, since this will save more
backhaul resources to serve the requests of the other data objects.
As can be seen from (\ref{eq:djk}), when the cache size $L_{C}$
is larger, the CoMP probability $1-\alpha$ is larger and more data
object requests can be handled by the cache-induced CoMP. Therefore,
the DoF stability region increases with the cache size $L_{C}$. Note
that even for the special case in Theorem \ref{thm:VIP-stability-region-closed},
the proposed online cache placement still has an advantage in terms
of the delay performance. In practice, the popularity varies over
time and different users have different preferences. Moreover, the
random user requests and wireless fading will also cause microscopic
spatial and temporary popularity variations. In this case, the online
cache placement can achieve much better delay performance (with the
same backhaul capacity $R$), as will be shown in simulations. 
\end{rem}

\subsection{Maximum Sum DoF under Identical User Popularity}

We shall derive a closed-form expression of the sum DoF under \textit{identical
user popularity}, which is a special case of identical user preference
when the arrival rate of any data object is the same for all users.
Specifically, the average arrival rates (in terms of DoF) have the
form $d_{j}^{k}=d\rho_{k},\forall j,k$, where $\rho_{k}$ can be
interpreted as the probability of requesting data object $k$, and
$d=\sum_{k\in\mathcal{K}}d_{j}^{k}$ is the total average arrival
rate of user $j$ (in terms of DoF). In this case, the maximum sum
DoF that can be achieved under the stability constraint is
\begin{equation}
D^{*}=\max_{d}Kd,\text{ s.t. }\left(d_{j}^{k}=d\rho_{k}\right)_{j\in\mathcal{U},k\in\mathcal{K}}\in\mathcal{D}_{ve}.\label{eq:exmsumDoF}
\end{equation}

\begin{thm}
[Maximum sum DoF under Zipf popularity]\label{thm:Maximum-sum-DoF}Consider
identical user popularity and suppose (\ref{eq:DoFaSM}) is satisfied.
When $r_{n}\geq ND^{A}$, the maximum sum DoF $D^{*}$ in (\ref{eq:exmsumDoF})
is
\begin{equation}
D^{*}=\begin{cases}
D_{A}, & R_{d}\geq R_{A}^{*}\\
\left(1-\alpha^{*}\right)D_{A}+\alpha^{*}D_{B}, & R_{d}\in\left(R_{B}^{*},R_{A}^{*}\right)\\
R_{d}/\sum_{k=L_{c}+1}^{K}\rho_{k}, & R_{d}\leq R_{B}^{*}
\end{cases},\label{eq:MAXDOF}
\end{equation}
where $D_{A}=\max_{d}d,\text{ s.t. },\left(d_{j}=d\right)_{j\in\mathcal{U}}\in\mathcal{D}^{A}$,
$D_{B}=\max_{d}d,\text{ s.t. },\left(d_{j}=d\right)_{j\in\mathcal{U}}\in\mathcal{D}^{B}$,
$R_{A}^{*}=ND_{A}\sum_{k=L_{c}+1}^{K}\rho_{k}$, $R_{B}^{*}=D_{B}\sum_{k=L_{c}+1}^{K}\rho_{k}$
and 
\begin{equation}
\alpha^{*}=\begin{cases}
\hat{\alpha}\triangleq\frac{R_{A}^{*}-NR_{d}}{R_{A}^{*}-NR_{B}^{*}} & \text{if }\frac{\left(1-\hat{\alpha}\right)R_{A}^{*}}{N}+\hat{\alpha}R_{B}^{*}\leq\hat{\alpha}D_{B}\\
\frac{R_{A}^{*}-R_{d}}{R_{A}^{*}-NR_{B}^{*}+\left(N-1\right)D_{B}} & \text{otherwise}.
\end{cases}\label{eq:Optalhpa}
\end{equation}
\end{thm}

Please refer to Appendix \ref{subsec:Proof-of-TheoremMDoF} for the
proof. The assumption $r_{n}\geq ND_{A}$ helps to simplify the expression
of the maximum sum DoF. This assumption is usually satisfied in practice
since the I/O speed of the storage device is typically much larger
than the wireless transmission rate. From Theorem \ref{thm:Maximum-sum-DoF},
we have the following observations.

\textbf{Impact of backhaul capacity:} When $R_{d}\geq R_{A}^{*}$,
there is enough backhaul capacity to support CoMP transmission with
probability 1. In this case, the sum DoF is $D_{A}$, which is completely
limited by the RAN. When $R_{d}\in\left(R_{B}^{*},R_{A}^{*}\right)$,
the backhaul capacity can only support CoMP transmission mode with
a non-zero probability less than 1. In this case, the sum DoF is between
$D_{B}$ and $D_{A}$, which is limited by both the RAN and backhaul.
Moreover, as $R_{d}$ increases from $R_{B}^{*}$ to $R_{A}^{*}$,
$\alpha^{*}$ decreases from 1 to 0, and $D^{*}$ increases from $D_{B}$
to $D_{A}$. When $R_{d}\leq R_{B}^{*}$, the sum DoF is less than
$D_{B}$, which is completely limited by the backhaul. 

\textbf{Impact of cache size $L_{C}$:} For a larger cache size $L_{C}$,
both $R_{A}^{*}$ and $R_{B}^{*}$ become smaller, i.e., a smaller
backhaul capacity is required to support full CoMP transmission. Moreover,
both the CoMP transmission probability $\alpha^{*}$ and the sum DoF
increase with $L_{C}$. On the other hand, when $L_{C}$ is very small,
the backhaul capacity has to be larger than $R_{A}^{*}\approx ND_{A}$
in order to support full CoMP transmission.

\textbf{Impact of popularity distribution: }As an example, consider
the \textit{Zipf popularity distribution} \cite{Yamakami_PDCAT06_Zipflaw},
where 
\begin{equation}
\rho_{k}=\frac{k^{-\varsigma}}{\sum_{k=1}^{K}k^{-\varsigma}},k=1,...,K,\label{eq:Zipf}
\end{equation}
and $\varsigma\geq0$ is the \textit{popularity skewness parameter}.
A larger popularity skewness $\varsigma$ means that the user requests
concentrate more on a few popular files. As a result, for a larger
$\varsigma$, both backhaul thresholds $R_{A}^{*}$ and $R_{B}^{*}$
become smaller. Moreover, both the CoMP transmission probability $\alpha^{*}$
and the sum DoF increase with $\varsigma$.

\section{Simulation Results\label{sec:Simulation-Results}}

Consider a cached MIMO interference network with seven BS-user pairs
placed in seven wrapped-around hexagonal cells. Each BS is equipped
with two antennas, and each user is equipped with one antenna. The
backhaul capacity per BS is 30 Mbps. The channel bandwidth is 10 MHz,
the slot size is 2 ms and the frame size is 0.5 s. The pathloss model
between BS $n$ and user $j$ is $PL_{j,n}=140.7+36.7\log10\left(d_{j,n}\right)$
\cite{3gpp_Rel9}, where $d_{j,n}$ is the distance between BS $n$
and user $j$. The channel between BS $n$ and user $j$ is modeled
as $\boldsymbol{H}_{jn}=PL_{j,n}\overline{\boldsymbol{H}}_{jn}$,
where $\overline{\boldsymbol{H}}_{jn}$ has i.i.d. Gaussian entries
of zero mean and unit variance. 

Zero-forcing beamforming (ZFBF), which is a special case of linear
precoding, is used at the PHY for both CoMP and coordinated transmission
modes. In the CoMP mode, all users can be simultaneously served by
the BSs, and the corresponding ZFBF precoder is given by $\boldsymbol{V}_{j}^{A}=\xi\widetilde{\boldsymbol{H}}^{H}\left(\widetilde{\boldsymbol{H}}\widetilde{\boldsymbol{H}}^{H}\right)^{-1}$,
where $\widetilde{\boldsymbol{H}}=\left[\widetilde{\boldsymbol{H}}_{j}\right]_{j=1,...,N}^{H}\in\mathbb{C}^{N\times2N}$
is the composite channel matrix between all BSs and all users; and
$\xi$ is chosen to satisfy the power constraint. In the coordinated
mode, we randomly select a subset of two users $\mathcal{U}^{B}$
for transmission at each time slot. For given user selection $\mathcal{U}^{B}$,
the corresponding ZFBF precoder is given by $\boldsymbol{V}_{j}^{B}=\sqrt{P}\overline{\boldsymbol{V}}_{j}^{B}$,
where $\overline{\boldsymbol{V}}_{j}^{B}\in\mathbb{C}^{2}$ with $\left\Vert \overline{\boldsymbol{V}}_{j}^{B}\right\Vert =1$
is obtained by the projection of $\boldsymbol{H}_{jn_{j}}$ on the
orthogonal complement of the subspace spanned by $\left[\boldsymbol{H}_{j^{'}n_{j}}\right]_{j^{'}\in\mathcal{U}^{B}\backslash\left\{ j\right\} }$.

There are $K=1000$ data objects in the content server. The data chunk
size is 50 KB and the data object size is 1 MB. At each user, object
requests arrive according to a Poisson process with a total average
arrival rate of $\lambda$ Mbps. To verify the performance under both
spatial and temporary popularity variations, we assume user $j$ only
requests a subset $\mathcal{F}_{j}$ of $100$ data objects whose
indices are randomly generated. The average arrival rate of data object
$\mathcal{F}_{j}\left(k\right)\in\mathcal{F}_{j}$ at user $j$ is
$\lambda_{j}^{\mathcal{F}_{j}\left(k\right)}=\lambda\rho_{k}$, where
$\mathcal{F}_{j}\left(k\right)$ is the $k$-th data object in $\mathcal{F}_{j}$
and $\rho_{k}$'s follow the Zipf distribution in (\ref{eq:Zipf}).
The following baselines are considered.

\textbf{Baseline 1 (Offline caching with dual-mode PHY }\cite{liu2013mixed,liu2014cache}\textbf{):
}Each BS caches the most popular $L_{C}$ data objects in an offline
manner. Dual-mode PHY is employed at the RAN.

\textbf{Baseline 2 (LFU with dual-mode PHY):} In Least Frequently
Used (LFU) caching, the nodes record how often each data object has
been requested and choose to cache the new data object if it is more
frequently requested than the least frequently requested cached data
object (which is replaced). Dual-mode PHY is employed at the RAN.

\textbf{Baseline 3 (VIP caching with single-mode PHY }\cite{yeh2014vip}\textbf{):
}The cache placement is determined by the VIP framework in \cite{yeh2014vip}
and only coordinated mode is considered at the PHY. 

\begin{figure}
\centering{}\includegraphics[width=85mm]{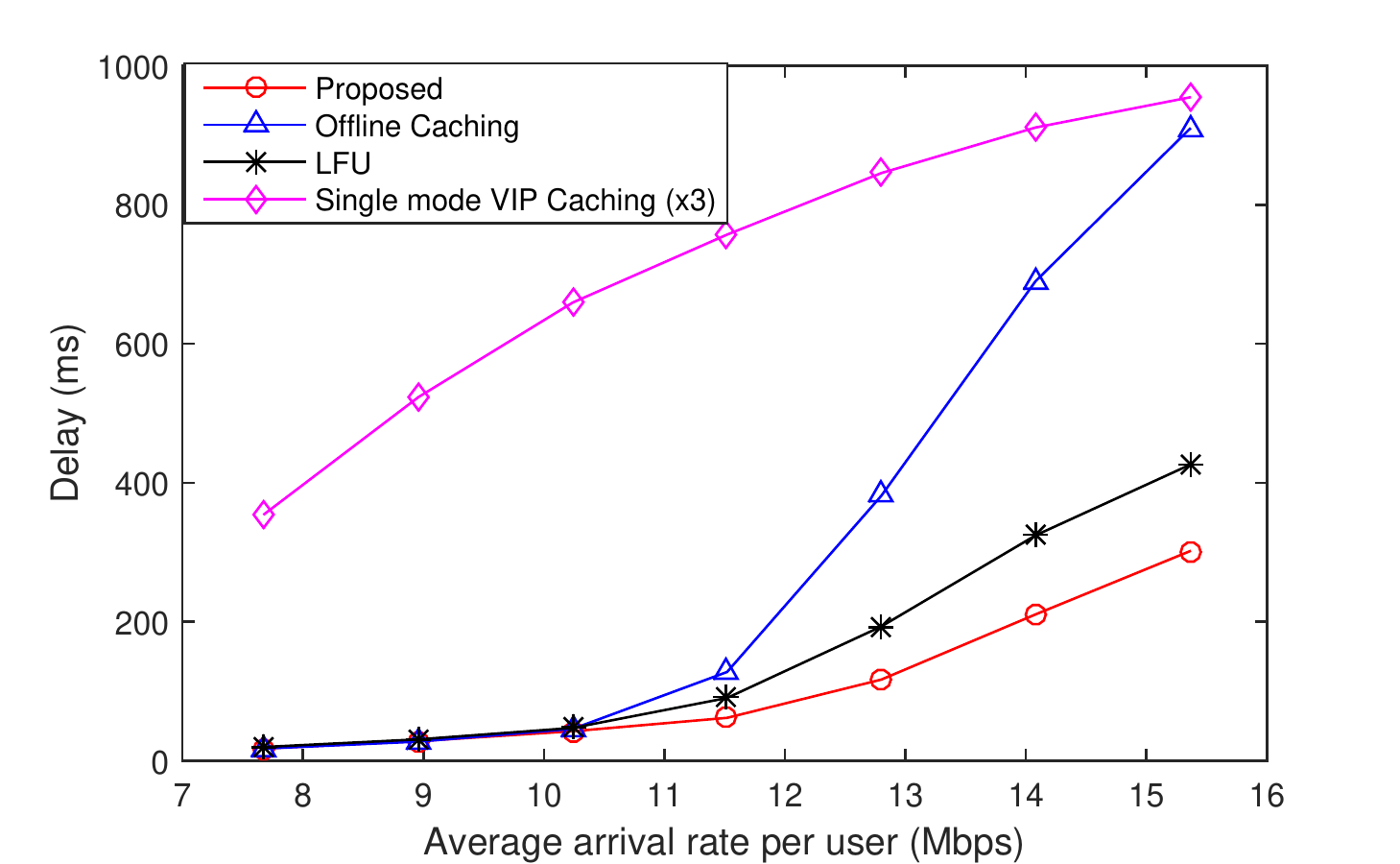}\caption{\label{fig:Darrival}{\small{}Delay versus per user average arrival
rate with cache size $L_{C}=80$ data objects and skewness $\varsigma=0.5$.}}
\end{figure}

\begin{figure}
\centering{}%
\begin{minipage}[t]{0.45\textwidth}%
\begin{center}
\includegraphics[width=85mm]{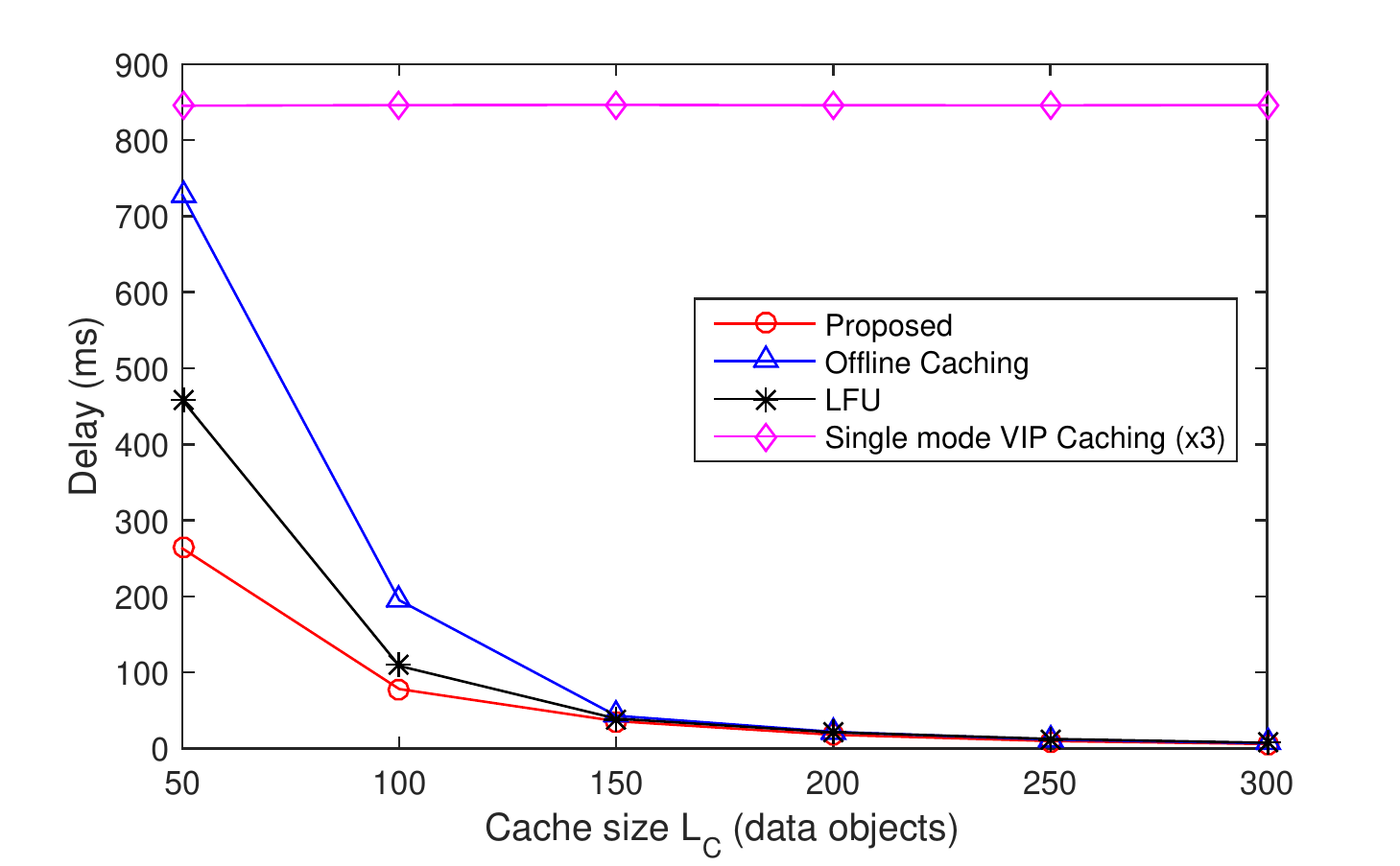}\caption{\label{fig:DLC}{\small{}Delay versus cache size $L_{C}$ with per
user average arrival rate $\lambda=13.25$ Mbps and skewness $\varsigma=0.5$.}}
\par\end{center}%
\end{minipage}\hfill{}%
\begin{minipage}[t]{0.45\textwidth}%
\begin{center}
\includegraphics[width=85mm]{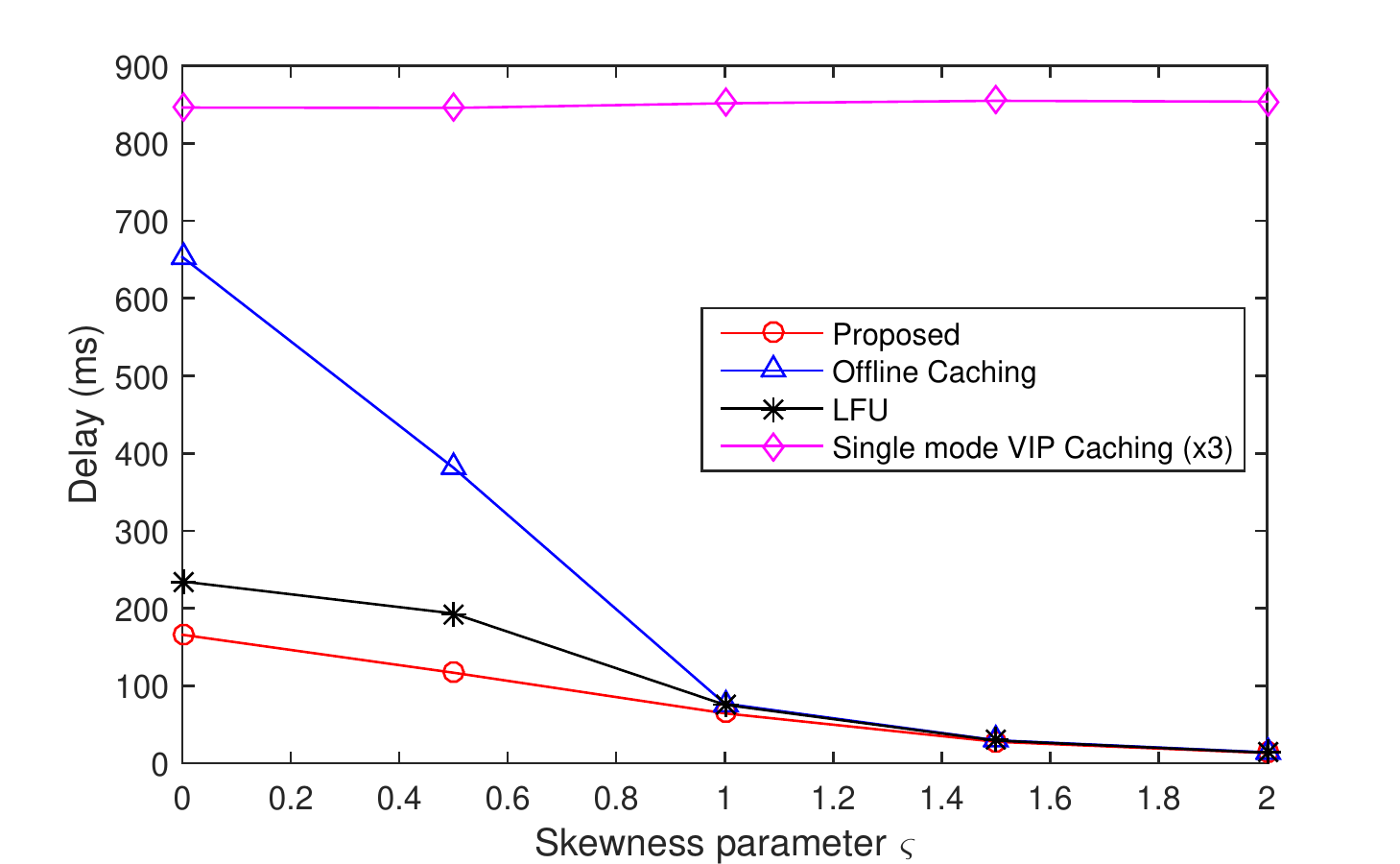}\caption{\label{fig:Dtau}{\small{}Delay versus skewness $\varsigma$ with
per user average arrival rate $\lambda=13.25$ Mbps and cache size
$L_{C}=80$ data objects.}}
\par\end{center}%
\end{minipage}
\end{figure}

For fair comparison, the data sub-channel $R_{d}$ and control sub-channel
$R_{c}$ are assumed to share the same $R$ Mbps backhaul capacity
for all schemes. In Fig. \ref{fig:Darrival} - \ref{fig:Dtau}, we
plot the delay performance of the schemes versus the average arrival
rate of each user $\lambda$, the cache size $L_{C}$ at each BS and
the skewness parameter $\varsigma$ respectively. For Baseline 3,
the delay shown in the figure is the actual delay divided by 3. The
delay for an IP request is the difference between the fulfillment
time (i.e., time of arrival of the requested DP) and the creation
time of the IP request. It can be seen that the delay of all schemes
increases with the average arrival rate $\lambda$, and decreases
with the cache size $L_{C}$ and skewness parameter $\varsigma$.
Moreover, the proposed scheme achieves better performance than all
baseline schemes. 

Note that when the cache size $L_{C}$ or skewness $\varsigma$ is
sufficiently large, the ``cache hit'' probability is high for any
caching scheme, and thus the performance gap between different caching
schemes will vanish, except for the single mode VIP caching scheme,
which still has a large performance gap w.r.t. the proposed scheme
because it cannot enjoy the cache-enabled opportunistic CoMP gain.
However, the proposed scheme has significant gain over all the baseline
schemes for practical scenarios when the cache size is limited compared
to the total content size and the popularity does not concentrate
on a few data objects. The fact that the proposed scheme achieves
a better performance than the LFU demonstrates the effectiveness of
the proposed mixed-timescale resource control algorithm based on the
dual-mode VIP and Lyapunov optimization framework. The LFU can achieve
a better performance than the offline caching scheme because it is
an online caching scheme which can better adapts the cached content
according to the microscopic spatial and temporary popularity variations.
Finally, the single mode VIP caching scheme is worse than the other
schemes because it cannot exploit the cache-enabled opportunistic
CoMP to enhance the capacity of RAN.

\section{Conclusion\label{sec:Conclusion}}

We propose a mixed-timescale online PHY caching and content delivery
scheme for wireless NDNs with a dual-mode PHY. The cache content placement
is performed once per frame ($T$ time slots) to avoid excessive cache
content placement cost. For a given cache state at each frame, the
PHY mode selection and rate allocation is performed once per time
slot to fully exploit the cached content at the BS. To facilitate
efficient resource control design, we introduce a \textit{dual-mode
VIP framework}, which transforms the original network into a virtual
network, and formulate the resource control design in the virtual
network. We establish the throughput optimality of the proposed solution.
Moreover, we obtain a closed-form expression for the maximum sum DoF
in the stability region under the Zipf popularity distribution. Simulations
show that the proposed solution outperforms the existing offline PHY
caching solution \cite{liu2013mixed,liu2014cache} and online caching
solutions \cite{yeh2014vip}.

\appendix

\subsection{Proof of Theorem \ref{thm:Equivalence-between-VA}\label{subsec:Proof-of-TheoremVAE}}

For a given $\boldsymbol{\lambda}\in\textrm{int}\Lambda_{v}$, the
minimum cost $\overline{\Gamma}_{v}^{*}\left(\boldsymbol{\lambda}\right)$
for VIP stability is given by the solution of the problem in Theorem
\ref{lem:VIP-stability-region}, minimized over the class of all stationary
randomized policies defined in Section \ref{subsec:VIP-Stability-Region}.
If $\boldsymbol{\lambda}\in\textrm{int}\Lambda_{v}$, there exists
a positive value $\epsilon$ such that $\boldsymbol{\lambda}+\boldsymbol{\epsilon}\in\textrm{int}\Lambda_{v}$.
It follows that there exists a stationary random resource control
policy $\left\{ p_{n}^{k}(i),M\left(t\right),\mu_{j}^{Ak}(t),\mu_{j}^{Bk}(t),\mu_{ng}^{k}(t)\right\} $
in the virtual network such that the corresponding conditional flow
balance constraint is satisfied:
\begin{align}
\lambda_{j}^{k}+\epsilon & \leq\overline{\mu}_{j|\boldsymbol{s}}^{Ak}+\overline{\mu}_{j|\boldsymbol{s}}^{Bk},\nonumber \\
\sum_{j\in\mathcal{U}}\overline{\mu}_{j|\boldsymbol{s}}^{Ak}+\overline{\mu}_{j_{n}|\boldsymbol{s}}^{Bk} & \leq\overline{\mu}_{ng|\boldsymbol{s}}^{k}+r_{n}s_{n}^{k}.\label{eq:CondVlemma}
\end{align}
By modifying the rate allocation policy to $\mu_{j}^{Ak}(t)=\mu_{j}^{Ak}(t)-\frac{0.5\epsilon}{N+1}$
and $\mu_{j}^{Bk}(t)=\mu_{j}^{Bk}(t)-\frac{0.5\epsilon}{N+1}$, the
resulting control policy satisfies the following conditional flow
balance constraint:
\begin{align}
\lambda_{j}^{k}+\frac{N\epsilon}{N+1} & \leq\overline{\mu}_{j|\boldsymbol{s}}^{Ak}+\overline{\mu}_{j|\boldsymbol{s}}^{Bk},\nonumber \\
\sum_{j\in\mathcal{U}}\overline{\mu}_{j|\boldsymbol{s}}^{Ak}+\overline{\mu}_{j_{n}|\boldsymbol{s}}^{Bk}+\frac{\epsilon}{2} & \leq\overline{\mu}_{ng|\boldsymbol{s}}^{k}+r_{n}s_{n}^{k},\label{eq:CondVlemma-1}
\end{align}
and we define $\overline{\Gamma}^{*}\left(\epsilon\right)$ as the
minimum average cost consumed by any such stationary policy.

In the following, we construct a random control policy $\left\{ p_{n}^{k}(i),m_{j}^{k}\left(t\right),c_{ng}(t),M_{a}\left(t\right),\boldsymbol{c}^{A}\left(t\right),\boldsymbol{c}^{B}\left(t\right)\right\} $
in the actual network such that the following conditional flow balance
constraint is satisfied when the arrival rate tuple is $\boldsymbol{\lambda}$:
\begin{align}
\overline{b}_{n|\boldsymbol{s}}+\frac{K\epsilon}{2} & \leq\overline{c}_{ng|\boldsymbol{s}},\overline{b}_{nj_{n}|\boldsymbol{s}}^{B}+\frac{KN\epsilon}{2\left(N+1\right)}\leq\overline{c}_{j_{n}|\boldsymbol{s}}^{B},\nonumber \\
 & \overline{b}_{nj|\boldsymbol{s}}^{A}+\frac{KN\epsilon}{2\left(N+1\right)}\leq\overline{c}_{j|\boldsymbol{s}}^{A}.\label{eq:condfloworg-1}
\end{align}
Specifically, the cache placement control policy $\left\{ p_{n}^{k}(i)\right\} $
in the actual network is the same as that in the virtual network.
For a given cache state $\boldsymbol{s}$, the forwarding mode at
time slot $t$ in the actual plane is randomly chosen with $\Pr\left[m_{j}^{k}\left(t\right)=1\right]=\min\left(\left(\overline{\mu}_{j|\boldsymbol{s}}^{Ak}-\frac{N\epsilon}{2\left(N+1\right)}\right)/\lambda_{j}^{k},1\right)$.
For the other control actions in the actual plane, we let $c_{ng}(t)=\sum_{k\in\mathcal{K}}\mu_{ng}^{k}(t)$,
$M_{a}\left(t\right)=M\left(t\right)$, $\boldsymbol{c}^{A}\left(t\right)=\boldsymbol{\mu}^{A}\left(t\right)$
and $\boldsymbol{c}^{B}\left(t\right)=\boldsymbol{\mu}^{B}\left(t\right)$.
Then, it can be verified that (\ref{eq:condfloworg-1}) is satisfied
and the above control policy achieves the same average cost $\overline{\Gamma}^{*}\left(\epsilon\right)$
as that in the virtual network. (\ref{eq:condfloworg-1}) implies
that the average departure rate of each DP queue is strictly larger
than the average arrival rate, and hence the network is stable \cite{neely2006energy}.
Therefore, we have proved that $\Lambda_{v}\subseteq\Lambda_{c}$.
Moreover, following a similar argument to Footnote 3 in \cite{neely2006energy},
it can be shown that $\overline{\Gamma}^{*}\left(\epsilon\right)\rightarrow\overline{\Gamma}_{v}^{*}\left(\boldsymbol{\lambda}\right)$
as $\epsilon\rightarrow0$, so that stability in the actual network
can be attained with an average cost that is arbitrarily close to
$\overline{\Gamma}_{v}^{*}\left(\boldsymbol{\lambda}\right)$.

Similarly, it can be shown that for any $\boldsymbol{\lambda}\in\textrm{int}\Lambda_{c}$,
the VIP stability in the virtual network can be attained with average
cost that is arbitrarily close to $\overline{\Gamma}_{c}^{*}\left(\boldsymbol{\lambda}\right)$.
This implies that $\Lambda_{c}\subseteq\Lambda_{v}$. Therefore, we
have $\Lambda_{c}=\Lambda_{v}$ and $\overline{\Gamma}_{c}^{*}\left(\boldsymbol{\lambda}\right)=\overline{\Gamma}_{v}^{*}\left(\boldsymbol{\lambda}\right)$
for any arrival rate tuple $\boldsymbol{\lambda}\in\textrm{int}\Lambda_{v}$.

\subsection{Proof of Theorem \ref{thm:Throughput-optimality-forVN}\label{subsec:Proof-of-TheoremThpopt}}

Let $\Omega_{R}^{*}$ denote the optimal random policy (the optimal
solution of (\ref{eq:minOmegaR})) in Lemma \ref{lem:VIP-stability-region}.
Let $\Delta_{T}^{\text{R}}(i)$ denote the T-step drift-plus-penalty
under $\Omega_{R}^{*}$.We first obtain an upper bound of $\Delta_{T}^{\text{R}}(i)$.
\begin{lem}
\label{lem: DBrandom}$\Delta_{T}^{\text{R}}(i)\leq\widetilde{\Delta}_{T}^{\text{R}}(i)=T^{2}B_{0}-T\epsilon\sum_{j\in N,k\in\mathcal{K}}V_{j}^{k}(t_{0})+TW\overline{\Gamma}_{v}^{*}\left(\boldsymbol{\lambda}\right)$,
where $B_{0}$ is a constant depending on $\left\{ \mu_{j,\text{max }}^{out},\mu_{n,\text{max }}^{out},\mu_{n,\text{max }}^{in}\right\} $
and $A_{\text{max}}$. 
\end{lem}
\begin{IEEEproof}
Following a similar analysis to the proof of Theorem 3 in \cite{neely2005dynamic},
it can be shown that
\[
\Delta_{T}^{\text{R}}(i)\leq\widetilde{\Delta}_{T}^{\text{R}}(i)=T^{2}B_{0}+\Delta_{T1}^{\text{R}}(i)+\Delta_{T2}^{\text{R}}(i)+TW\overline{\Gamma}_{v}^{*}\left(\boldsymbol{\lambda}\right),
\]
\begin{align*}
\widetilde{\Delta}_{T1}^{\text{R}}(i) & =-2T\sum_{j\in\mathcal{U},k\in\mathcal{K}}V_{j}^{k}(t_{0}^{i})\mathbb{E}\bigg[\frac{1}{T}\sum_{\tau=t_{0}}^{t_{0}+T-1}\bigg(M^{*}\left(t\right)\sum_{j\in\mathcal{U}}\mu_{j}^{Ak*}(t)\\
 & +\overline{M}^{*}\left(t\right)\sum_{j\in\mathcal{U}}\mu_{j}^{Bk*}(t)\bigg)-\lambda_{n}^{k}|\boldsymbol{s}(i-1)\bigg],
\end{align*}
\begin{align*}
\widetilde{\Delta}_{T2}^{\text{R}}(i) & =-2T\sum_{j\in\mathcal{B},k\in\mathcal{K}}V_{j}^{k}(t_{0})\mathbb{E}\bigg[-\frac{1}{T}\sum_{\tau=t_{0}}^{t_{0}+T-1}\bigg(M^{*}\left(\tau\right)\sum_{j\in\mathcal{U}}\mu_{j}^{Ak*}(\tau)+\overline{M}^{*}\left(\tau\right)\sum_{j\in\mathcal{U}}\mu_{j}^{Bk*}(\tau)\bigg)\\
 & +\frac{1}{T}\sum_{\tau=t_{0}}^{t_{0}+T-1}\mu_{jg}^{k}(\tau)^{*}+r_{n}\left[s_{n}^{k}(t_{0}^{i}-1)+p_{n}^{k}(t_{0})^{*}\right]|\boldsymbol{s}(i-1)\bigg],
\end{align*}
where the control actions with superscript $^{*}$ are given by the
optimal random policy $\Omega_{R}^{*}$. Since for any given $\boldsymbol{s}(i)$,
$\Omega_{R}^{*}$ satisfies the conditional flow balance (\ref{eq:condfloworgV}),
we have $\widetilde{\Delta}_{T1}^{\text{R}}(i)\leq-2T\sum_{j\in\mathcal{U},k\in\mathcal{K}}V_{j}^{k}(t_{0}^{i})\epsilon$
and $\widetilde{\Delta}_{T2}^{\text{R}}(i)\leq-2T\sum_{j\in\mathcal{B},k\in\mathcal{K}}V_{j}^{k}(t_{0}^{i})\epsilon$,
from which Lemma \ref{lem: DBrandom} follows.
\end{IEEEproof}
To obtain an upper bound of the drift for the proposed solution, we
need to consider a FRAME policy which serves as a bridge to connect
the proposed solution and the optimal random policy $\Omega_{R}^{*}$.
In the FRAME policy, the cache content placement is the same as the
proposed solution, while the mode selection and rate allocation is
the optimal solution of a modified version of the drift minimization
problem in (\ref{eq:1stepmin}), with the current VIP length $\left\{ V_{j}^{k}(t),V_{n}^{k}(t)\right\} $
replaced by the outdated VIP length $\left\{ V_{j}^{k}(t_{0}^{i}),V_{n}^{k}(t_{0}^{i})\right\} $.
Let $\widetilde{\Delta}_{T}^{\textrm{P}}(i)$ and $\widetilde{\Delta}_{T}^{\textrm{F}}(i)$
denote the upper bound of the T-step drift-plus-penalty defined in
Theorem \ref{thm: T-step-Drift-Plus-Penalty-Upper} under the proposed
solution and the FRAME policy respectively. The following lemma states
the relationship between the different policies.
\begin{lem}
\label{lem: driftrelation}$\widetilde{\Delta}_{T}^{\textrm{P}}(i)-\widetilde{\Delta}_{T}^{\textrm{F}}(i)\leq T^{2}B_{1}$,
where $B_{1}$ is a constant depending on $\mu_{n,\text{max }}^{out}$,
$\mu_{n,\text{max }}^{in}$ and $A_{\text{max}}$. Moreover, $\widetilde{\Delta}_{T}^{\text{F}}(i)\leq\widetilde{\Delta}_{T}^{\text{R}}(i)$.
\end{lem}
\begin{IEEEproof}
Similar to the proof of Lemma 5 in \cite{neely2005dynamic}, the result
$\widetilde{\Delta}_{T}^{\textrm{P}}(i)-\widetilde{\Delta}_{T}^{\textrm{F}}(i)\leq T^{2}B_{1}$
follows from the fact that the expected magnitude of change in a single
VIP queue is at most $(\mu_{j,\text{max }}^{out}+A_{\text{max}}),\forall j\in\mathcal{U}$
and $(\mu_{n,\text{max }}^{out}+\mu_{n,\text{max }}^{in}+r_{n}),\forall n\in\mathcal{B}$,
and the caching-related term in $\widetilde{\Delta}_{T}^{\textrm{P}}(i)$
and $\widetilde{\Delta}_{T}^{\textrm{F}}(i)$ is identical. The detailed
proof is omitted for conciseness. On the other hand, the result $\widetilde{\Delta}_{T}^{\text{F}}(i)\leq\widetilde{\Delta}_{T}^{\text{R}}(i)$
follows from the fact that the FRAME policy minimizes $\widetilde{\Delta}_{T}\left(i\right)$.
\end{IEEEproof}
From Lemma \ref{lem: DBrandom} and \ref{lem: driftrelation}, we
conclude that 
\begin{align}
\Delta_{T}^{\textrm{P}}(i) & \leq\widetilde{\Delta}_{T}^{\textrm{P}}(i)\leq\widetilde{\Delta}_{T}^{\text{F}}(i)+T^{2}B_{1}\nonumber \\
 & \leq T^{2}B-T\epsilon\sum_{j\in N,k\in\mathcal{K}}V_{j}^{k}(t_{0})+TW\overline{\Gamma}_{v}^{*}\left(\boldsymbol{\lambda}\right),\label{eq:BoundP}
\end{align}
where $B\triangleq B_{0}+B_{1}$. Finally, Theorem \ref{thm:Throughput-optimality-forVN}
can be proved from (\ref{eq:BoundP}) by applying Theorem 4.2 of \cite{neely2010stochastic}
and the same technique as in Theorem 3 of \cite{neely2005dynamic}.

\subsection{Proof of Theorem \ref{thm:VIP-stability-region-closed}\label{subsec:VIP-Stability-Region} }

Consider a \textit{stationary random policy} $\Omega_{R}$ as follows.
At the beginning of the $i$-th frame, $\left\{ p_{n}^{k}(i)\right\} $
is randomly chosen from the set of all feasible cache placement control
actions $\mathcal{A}_{p}^{\boldsymbol{s}^{'}}$ (i.e., satisfying
the cache size constraint (\ref{eq:cachesizecon})) with probability
$\boldsymbol{\alpha}_{p}^{\boldsymbol{s}^{'}}=\left[\alpha_{p1}^{\boldsymbol{s}^{'}},...,\alpha_{p\left|\mathcal{A}_{p}^{\boldsymbol{s}^{'}}\right|}^{\boldsymbol{s}^{'}}\right]$,
where $\boldsymbol{s}^{'}=\boldsymbol{s}\left(i-1\right)$. Here,
we use the superscript $\boldsymbol{s}^{'}$ to indicate that $\mathcal{A}_{p}^{\boldsymbol{s}^{'}}$
and $\boldsymbol{\alpha}_{p}^{\boldsymbol{s}^{'}}$ depend on $\boldsymbol{s}^{'}$.
At each time slot $t$, $M\left(t\right)$ is randomly chosen with
$\Pr\left[M\left(t\right)=0\right]=\alpha_{M}^{\left(\boldsymbol{s},\boldsymbol{H}\right)}$
and $\Pr\left[M\left(t\right)=1\right]=1-\alpha_{M}^{\left(\boldsymbol{s},\boldsymbol{H}\right)}$,
where $\boldsymbol{s}=\boldsymbol{s}\left(i\right)$ and $\boldsymbol{H}=\boldsymbol{H}\left(t\right)$;
$\mu_{j}^{Ak}\left(t\right)$ is randomly chosen from $N+1$ rate
tuples $\left\{ \mu_{ji}^{Ak\left(\boldsymbol{s},\boldsymbol{H}\right)},i=1,...,N+1\right\} $
satisfying
\begin{equation}
\left[\sum_{k\in\mathcal{K}}\mu_{ji}^{Ak\left(\boldsymbol{s},\boldsymbol{H}\right)}\right]_{j\in\mathcal{U}}\in C^{A}\left(\boldsymbol{H}\right),i=1,...,N+1,\label{eq:RdC1con}
\end{equation}
with probability $\boldsymbol{\alpha}_{A}^{\left(\boldsymbol{s},\boldsymbol{H}\right)}=\left[\alpha_{A1}^{\left(\boldsymbol{s},\boldsymbol{H}\right)},...,\alpha_{A\left(N+1\right)}^{\left(\boldsymbol{s},\boldsymbol{H}\right)}\right]$;
$\mu_{j}^{Bk}\left(t\right)$ is randomly chosen from $N+1$ rate
tuples $\left\{ \mu_{ji}^{Bk\left(\boldsymbol{s},\boldsymbol{H}\right)},i=1,...,N+1\right\} $
satisfying 
\begin{equation}
\left[\sum_{k\in\mathcal{K}}\mu_{ji}^{Bk\left(\boldsymbol{s},\boldsymbol{H}\right)}\right]_{j\in\mathcal{U}}\in C^{B}\left(\boldsymbol{H}\right),i=1,...,N+1,\label{RdC0con}
\end{equation}
with probability $\boldsymbol{\alpha}_{B}^{\left(\boldsymbol{s},\boldsymbol{H}\right)}=\left[\alpha_{B1}^{\left(\boldsymbol{s},\boldsymbol{H}\right)},...,\alpha_{B\left(N+1\right)}^{\left(\boldsymbol{s},\boldsymbol{H}\right)}\right]$;
and $\mu_{ng}^{k}(t)=\mu_{ng}^{k\left(\boldsymbol{s},\boldsymbol{H}\right)}$
satisfying 
\begin{equation}
\sum_{k\in\mathcal{K}}\mu_{ng}^{k\left(\boldsymbol{s},\boldsymbol{H}\right)}\leq R_{d}.\label{eq:Rdbkcon}
\end{equation}
Moreover, the above random policy {\small{}$\Omega_{R}=\left\{ \mathcal{A}_{p}^{\boldsymbol{s}^{'}},\boldsymbol{\alpha}_{p}^{\boldsymbol{s}^{'}},\alpha_{M}^{\left(\boldsymbol{s},\boldsymbol{H}\right)},\mu_{ji}^{Ak\left(\boldsymbol{s},\boldsymbol{H}\right)},\boldsymbol{\alpha}_{A}^{\left(\boldsymbol{s},\boldsymbol{H}\right)},\mu_{ji}^{Bk\left(\boldsymbol{s},\boldsymbol{H}\right)},\boldsymbol{\alpha}_{B}^{\left(\boldsymbol{s},\boldsymbol{H}\right)},\mu_{ng}^{k\left(\boldsymbol{s},\boldsymbol{H}\right)}\right\} $
}is called a stationary random policy if the resulting Markov cache
state process $\boldsymbol{s}\left(i\right)$ is stationary. Let $P_{\boldsymbol{s}^{'},\boldsymbol{s}}^{\Omega_{R}}$
denote the transition probability of the controlled Markov Process
$\boldsymbol{s}\left(i\right)$ from state $\boldsymbol{s}^{'}$ to
state $\boldsymbol{s}$, and $\pi_{\boldsymbol{s}}^{\Omega_{R}}$
denote the steady state probability of $\boldsymbol{s}\left(i\right)=\boldsymbol{s}$,
under the stationary random policy $\Omega_{R}$. Then we have the
following lemma.
\begin{lem}
[VIP stability region]\label{lem:VIP-stability-region}The VIP stability
region $\Lambda_{v}$ consists of all arrival rate tuples $\boldsymbol{\lambda}$
such that there exists a stationary random policy $\Omega_{R}$ satisfying
(\ref{eq:RdC1con}), (\ref{RdC0con}), (\ref{eq:Rdbkcon}) and 
\begin{align}
 & \lambda_{j}^{k}\leq\mathbb{E}_{\boldsymbol{H}}\bigg[\left(1-\alpha_{M}^{\left(\boldsymbol{s},\boldsymbol{H}\right)}\right)\sum_{i=1}^{N+1}\alpha_{Ai}^{\left(\boldsymbol{s},\boldsymbol{H}\right)}\mu_{ji}^{Ak\left(\boldsymbol{s},\boldsymbol{H}\right)}+\alpha_{M}^{\left(\boldsymbol{s},\boldsymbol{H}\right)}\sum_{i=1}^{N+1}\alpha_{Bi}^{\left(\boldsymbol{s},\boldsymbol{H}\right)}\mu_{ji}^{Bk\left(\boldsymbol{s},\boldsymbol{H}\right)}\bigg],\nonumber \\
 & \mathbb{E}_{\boldsymbol{H}}\bigg[\left(1-\alpha_{M}^{\left(\boldsymbol{s},\boldsymbol{H}\right)}\right)\sum_{j\in\mathcal{U}}\sum_{i=1}^{N+1}\alpha_{Ai}^{\left(\boldsymbol{s},\boldsymbol{H}\right)}\mu_{ji}^{Ak\left(\boldsymbol{s},\boldsymbol{H}\right)}+\alpha_{M}^{\left(\boldsymbol{s},\boldsymbol{H}\right)}\sum_{i=1}^{N+1}\alpha_{Bi}^{\left(\boldsymbol{s},\boldsymbol{H}\right)}\mu_{j_{n}i}^{Bk\left(\boldsymbol{s},\boldsymbol{H}\right)}\bigg]\leq\mathbb{E}_{\boldsymbol{H}}\left[\mu_{ng}^{k\left(\boldsymbol{s},\boldsymbol{H}\right)}\right]+r_{n}s_{n}^{k},\label{eq:RdCondflow}
\end{align}
$\forall j\in\mathcal{U},k\in\mathcal{K},n\in\mathcal{B}$ and $\forall\boldsymbol{s}\in\mathcal{S}$.
Moreover, for any $\boldsymbol{\lambda}\in\textrm{int}\Lambda_{v}$,
the minimum cache content placement cost required for stability $\overline{\Gamma}_{v}^{*}\left(\boldsymbol{\lambda}\right)$
is given by
\begin{align}
\overline{\Gamma}_{v}^{*}\left(\boldsymbol{\lambda}\right)=\min_{\Omega_{R}} & \sum_{\left(\boldsymbol{s}^{'},\boldsymbol{s}\right)\in\mathcal{S}_{T}}\pi_{\boldsymbol{s}^{'}}^{\Omega_{R}}P_{\boldsymbol{s}^{'},\boldsymbol{s}}^{\Omega_{R}}\sum_{n\in\mathcal{B},k\in\mathcal{K}}\boldsymbol{1}_{\left\{ s_{n}^{k}-s_{n}^{'k}=1\right\} }\label{eq:minOmegaR}\\
\text{s.t. } & (\ref{eq:RdC1con}),\:(\ref{RdC0con}),\:(\ref{eq:Rdbkcon})\text{ and }(\ref{eq:RdCondflow})\text{ are satisfied}\nonumber \\
 & \forall j\in\mathcal{U},k\in\mathcal{K},n\in\mathcal{B},\boldsymbol{s}\in\mathcal{S}.\nonumber 
\end{align}
\end{lem}

\begin{IEEEproof}
The proof involves showing that $\boldsymbol{\lambda}\in\Lambda_{v}$
is necessary for stability and that $\boldsymbol{\lambda}\in\textrm{int}\Lambda_{v}$
is sufficient for stability. First, we show $\boldsymbol{\lambda}\in\Lambda_{v}$
is necessary for stability. We take the arrival and departure at the
base stations for example and the same technique can be applied to
the virtual queues at the end users. By Lemma 1 of \cite{neely2005dynamic},
network stability implies there exists a finite $Z$ such that $V_{n}^{k}(t)\leq Z$
for all $n\in\mathcal{B}$ and $k\in\mathcal{K}$ holds infinitely
often. Given an arbitrarily small value $\epsilon>0$, there exists
a slot $\widetilde{t}$ such that 
\begin{equation}
V_{n}^{k}(\widetilde{t})\leq Z,\quad\frac{Z}{\widetilde{t}}\leq\epsilon.\label{eq:largeZbound}
\end{equation}
Assuming $V_{n}^{k}(t_{0})=0$, we have

\begin{equation}
-V_{n}^{k}(t)\leq\sum_{\tau=t_{0}}^{t}F_{ng}^{k(\mathbf{s}(i),\boldsymbol{H}(\tau))}(\tau)+r_{n}\sum_{t=t_{0}}^{t}S_{n}^{k}(\tau)-\sum_{\tau=t_{0}}^{t}M(\tau)\sum_{j\in\mathcal{U}}F_{ji}^{Ak(\mathbf{s}(i),\boldsymbol{H}(\tau))}-\sum_{\tau=t_{0}}^{t}\overline{M}(\tau)\sum_{j\in\mathcal{U}}F_{j_{n}i}^{Bk(\mathbf{s}(i),\boldsymbol{H}(\tau))},\label{eq:virtualQueuePackets}
\end{equation}
 where $F_{ng}^{k(\mathbf{s}(i),\boldsymbol{H}(\tau))}(\tau)$ represents
the actual virtual packet transmited through the backhual for content
$k$ at time $\tau$. Thus, by \ref{eq:largeZbound} and \ref{eq:virtualQueuePackets},
we have

\begin{equation}
0\leq\epsilon+\frac{1}{\widetilde{t}}\sum_{\tau=t_{0}}^{\widetilde{t}}F_{ng}^{k(\mathbf{s}(i),\boldsymbol{H}(\tau))}(\tau)+r_{n}\frac{1}{\widetilde{t}}\sum_{\tau=t_{0}}^{\widetilde{t}}S_{n}^{k}(\tau)-\frac{1}{\widetilde{t}}\sum_{\tau=t_{0}}^{\widetilde{t}}M(\tau)\sum_{j\in\mathcal{U}}F_{ji}^{Ak(\mathbf{s}(i),\boldsymbol{H}(\tau))}-\frac{1}{\widetilde{t}}\sum_{\tau=t_{0}}^{\widetilde{t}}\overline{M}(\tau)\sum_{j\in\mathcal{U}}F_{j_{n}i}^{Bk(\mathbf{s}(i),\boldsymbol{H}(\tau))}.\label{eq:multiStepFlowBlance}
\end{equation}
Note that the VIP stability region is defined under conditional flow
balance. Given \ref{eq:multiStepFlowBlance}, it remains to prove
that the constructed stationary random policy $\Omega_{R}$ satisfies
the posed requirements. By letting $\epsilon\rightarrow0$ or $\frac{1}{\widetilde{t}}\rightarrow\infty$,
the total virutal packets drained by cache $r_{n}\frac{1}{\widetilde{t}}\sum_{\tau=t_{0}}^{\widetilde{t}}S_{n}^{k}(\tau)$
converges to its steady state distribution $r_{n}s_{n}^{k}$. For
the rate allocatation terms such as $\frac{1}{\widetilde{t}}\sum_{\tau=t_{0}}^{\widetilde{t}}M(\tau)\sum_{j\in\mathcal{U}}F_{ji}^{Ak(\mathbf{s}(i),\boldsymbol{H}(\tau))}$,
since the evolution of the cache state and channel state is independent
of the rate allocation decision, and for each given $\mathbf{s}$,
$\frac{1}{\widetilde{t}}\sum_{\tau=t_{0}}^{\widetilde{t}}M(\tau)\sum_{j\in\mathcal{U}}F_{ji}^{Ak(\mathbf{s},\boldsymbol{H}(\tau))}\rightarrow\mathbb{E}_{\boldsymbol{H}}\bigg[\left(1-\alpha_{M}^{\left(\boldsymbol{s},\boldsymbol{H}\right)}\right)\sum_{j\in\mathcal{U}}\sum_{i=1}^{N+1}\alpha_{Ai}^{\left(\boldsymbol{s},\boldsymbol{H}\right)}\mu_{ji}^{Ak\left(\boldsymbol{s},\boldsymbol{H}\right)}\bigg]$.
By applying the steady state distribution of $\mathbf{s}$, \ref{eq:RdCondflow}
follows. Next, we show $\boldsymbol{\lambda}\in\textrm{int}\Lambda_{v}$
is sufficient for stability. By apply the condition \ref{eq:RdCondflow},
we have for each virtual queue, the arrival rate is less than the
service rate, so the network is stable. 
\end{IEEEproof}
Finally, Theorem \ref{thm:VIP-stability-region-closed} follows from
Lemma \ref{lem:VIP-stability-region} and the definition of $\mathcal{D}_{ve}$.

\subsection{Proof of Theorem \ref{thm:Maximum-sum-DoF}\label{subsec:Proof-of-TheoremMDoF}}

It follows from the symmetry property of the problem that $\mu_{j}^{Ak}=\mu^{Ak},\forall j$
and $\mu_{j}^{Bk}=\mu^{Bk},\forall j$ at the optimal solution of
(\ref{eq:exmsumDoF}). As a result, when $r_{n}\geq ND^{A}$, Problem
(\ref{eq:exmsumDoF}) is equivalent to the following problem:
\begin{align*}
 & \max_{d,\alpha,\mu^{Ak},\mu^{Bk}}Kd,\text{ s.t. }d\rho_{k}=\left(1-\alpha\right)\mu^{Ak}+\alpha\mu^{Bk},\forall k\\
 & N\left(1-\alpha\right)\sum_{k=L_{c}+1}^{K}\mu^{Ak}+\alpha\sum_{k=L_{c}+1}^{K}\mu^{Bk}\leq R_{d},\\
 & \sum_{k\in\mathcal{K}}\mu^{Ak}\leq D_{A},\:\sum_{k\in\mathcal{K}}\mu^{Bk}\leq D_{B}.
\end{align*}
By finding the optimal solution of the above problem, we can obtain
the maximum sum DoF as in Theorem \ref{thm:Maximum-sum-DoF}.


\end{document}